\documentclass[
 reprint,
superscriptaddress,
 amsmath,amssymb,
 aps,
pre,
twocolumn]{revtex4-2}

\usepackage{graphicx}
\usepackage{dcolumn}
\usepackage{bm}
\usepackage{xcolor}
\usepackage{subcaption}
\usepackage{multirow}
\usepackage{booktabs}
\usepackage{array}
\usepackage{float}
\usepackage{svg}

\begin{document}

\preprint{APS/123-QED}

\title{Community detection robustness of graph neural networks}

\author{Jaidev Goel}
\email[]{jaidev@vt.edu}
\thanks{corresponding author}
\affiliation{Department of Statistics, Virginia Polytechnic Institute and State University, Blacksburg, VA 24061, USA}
\author{Pablo Moriano}
\email[]{moriano@ornl.gov}
\homepage[]{https://pmoriano.com}
\affiliation{Computer Science and Mathematics Division, Oak Ridge National Laboratory, Oak Ridge, TN 37830, USA}
\author{Ramakrishnan Kannan}
\affiliation{Computer Science and Mathematics Division, Oak Ridge National Laboratory, Oak Ridge, TN 37830, USA}
\author{Yulia R. Gel}
\affiliation{Department of Statistics, Virginia Polytechnic Institute and State University, Blacksburg, VA 24061, USA}





 


\begin{abstract}


Graph neural networks (GNNs) are increasingly widely used for community detection in attributed networks. They combine structural topology with node attributes through message passing and pooling. However, their  robustness or lack of thereof with respect to different perturbations and targeted attacks in conjunction with community detection tasks is not well understood. To shed light into latent mechanisms behind GNN sensitivity on community detection tasks, we conduct a systematic computational evaluation 
of six widely adopted GNN architectures: GCN, GAT, GraphSAGE, DiffPool, MinCUT, and DMoN. The analysis covers three perturbation categories: node attribute manipulations, edge topology distortions, and adversarial attacks. We use element-centric similarity as the evaluation metric on synthetic benchmarks and real-world citation networks. Our findings indicate that supervised GNNs tend to achieve higher baseline accuracy, while unsupervised methods, particularly DMoN, maintain stronger resilience under targeted and adversarial perturbations. Furthermore, robustness appears to be strongly influenced by community strength, with well-defined communities reducing performance loss. Across all models, node attribute perturbations associated with targeted edge deletions and shift in attribute distributions tend to cause the largest degradation in community recovery. These findings highlight important trade-offs between accuracy and robustness in GNN-based community detection and offer new
insights into selecting architectures resilient to noise and adversarial attacks.
\end{abstract}
\maketitle

\section{\label{sec:level2}Introduction}

Community detection, sometimes termed clustering in network analysis, addresses the problem of identifying groups of nodes with dense intra-group connections and sparse inter-group links~\citep{fortunato2010community}. It plays a central role in uncovering the mesoscale structure of complex systems, with applications ranging from social network analysis and biological interaction mapping to infrastructure design and information retrieval~\citep{newman2006modularity,girvan2002community,wandelt2021estimation,lancichinetti2009community}. Accurate detection of communities provides crucial insights into the organization, dynamics, and functionality of these systems, enabling the identification of processes such as information diffusion~\cite{moriano2019community}, functional modularity~\citep{aref2024analyzing,schuurman2024modularity}, and collective behavior~\citep{khawaja2024exploring}.

Graph neural networks (GNNs) have emerged as powerful tools for network representation learning by integrating structural topology with node features through message-passing operations~\citep{scarselli2008graph,wu2020comprehensive,xu2018powerful,velivckovic2023everything}. Architectures such as graph convolutional networks (GCNs)~\citep{kipf2016semi}, graph attention networks (GATs)~\citep{velivckovic2017gat}, and GraphSAGE~\citep{hamilton2017sage} have achieved strong performance in node classification and graph-level prediction tasks. These methods have recently been adapted for community detection~\citep{su2022comprehensive_survey,ren2024deepsurvey,liu2022survey}. Advances in end-to-end differentiable clustering, such as DiffPool~\citep{ying2018hierarchical_diffpool}, MinCUT~\citep{bianchi2020spectral_mincut}, and DMoN~\citep{tsitsulin2023graph}, use hierarchical pooling and cluster quality optimization to improve the recovery of latent community structures.

Despite these advances, due to the structure of GNNs and their use of message-passing, small perturbations to edges or node attributes can propagate through the model. This sensitivity is particularly concerning for community detection, where the goal is recover meaningful structures in noisy real-world networks. Although robustness has been studied extensively for GNNs in classification tasks~\citep{nettack,mettack,wang2021certified,gunnemann2022graph,zhang2022graph,zhang2024trustworthy}, the stability of GNN-based community detection under graph perturbations remains largely unexplored. Real-world networks often contain noise, incomplete data, targeted disruptions, and adversarial manipulations~\citep{sun2022adversarial}. Existing robustness studies for community detection have focused mainly on classical algorithms under edge addition or removal~\citep{tian2023robustness,wei2024robustness}, with limited attention to modern GNN-based approaches. In addition, prior studies primarily examine isolated perturbation types, while the real world scenarios often involve multiple interacting sources of perturbation. Our goal is to fill this important gap in our understanding of the sensitivity of GNN-based community detection to various attacks and latent mechanisms behind it.

Let $\mathcal{G} = (V, E)$ denote a graph with adjacency matrix $\mathbf{A}$ and node features $\mathbf{X}$. A GNN-based community detection model $f: \mathcal{G} \rightarrow C$ outputs a partition $C$ of the nodes. Robustness analysis considers perturbed graphs $\mathcal{G}^\ast$ from an admissible set $\mathcal{P}$, constrained by budget of how dissimilar the perturbed graph can be to the original graph metric. For community detection task, robustness analysis then evaluates a similarity score $S(C, C^\ast)$ between original and perturbed partitions.

To the best of our knowledge, we present the first comprehensive robustness benchmark for GNN-based community detection. We evaluate six representative architectures, including three supervised models (GCN, GAT, GraphSAGE) and three unsupervised models (DiffPool, MinCUT, DMoN), across three types of perturbations: (1) node attribute manipulations, (2) structural perturbations from random and betweenness-centrality-based edge removals, and (3) adversarial attacks using Nettack and Metattack. Our experiments span synthetic Lancichinetti–Fortunato–Radicchi (LFR) graphs~\citep{lancichinetti2008benchmark}, the degree-corrected stochastic block model (DC-SBM)~\citep{karrer2011stochastic, dey2022community, zhao2012consistency,  tsitsulin2022synthetic}, and real-world citation networks. The results show that supervised GNNs tend to maintain higher robustness under mild random perturbations, while unsupervised models exhibit greater resilience to targeted and adversarial attacks. These findings reveal important distinct trade-offs in architecture choice depending on the perturbation regime. Robustness trends are quantified using element-centric similarity (ECS)~\citep{gates2019element}, which provides a detailed view of community structure preservation. Our findings suggest that while supervised GNN methods (GCN, GAT, GraphSAGE) achieve better baseline performance on clean data, modularity based unsupervised method DMoN demonstrates higher robustness under various perturbations, maintaining consistent community detection performance even under moderate patterns of perturbations. Community strength, perturbation type, and architectural design emerge as key factors influencing robustness, with both perturbations to attributes and graph topology being effective at affecting robustness of GNN based community detection. The observed phenomena opens new perspectives not only for more systematic guidance in selecting GNN models for community detection tasks but also for developing more robust task-tailored GNN architectures. The code is available to reproduce all the results at~\cite{github_code}.

\section{\label{sec:level2}Methodology}



\subsection{\label{sec:level2}Synthetic graphs}

\subsubsection{LFR benchmark}

The Lancichinetti--Fortunato--Radicchi (LFR) benchmark has emerged as a popular tool for generating graphs with known community structures~\cite{lancichinetti2008benchmark,lancichinetti2009benchmarks}. Its primary strength lies in its ability to create networks where both degree and community size distributions follow power-laws, mirroring the characteristics of many real-world networks. Key parameters in the LFR benchmark include $\alpha$ and $\beta$, which are exponents controlling the degree and community size distributions, respectively. Other important parameters are the average degree $\langle k \rangle$, maximum degree $\text{max}_k$, minimum and maximum community sizes $\text{min}_c$ and $\text{max}_c$, and the mixing parameter $\mu$. The mixing parameter $\mu$ is particularly significant, as it determines the fraction of inter-community edges. Lower $\mu$ values result in stronger community structures~\citep{fortunato2010community}. In our experiments, we use the {\texttt{NetworkX}} Python package \cite{hagberg2008exploring} to generate undirected, unweighted LFR benchmark graphs with 1,000 nodes. We vary the mixing parameter $\mu$ ranging between 0.1 to 0.5 to investigate the effects of perturbations at different initial community strengths. 


\subsubsection{Stochastic block model (SBM)}The standard (degree-free) SBM is a probabilistic framework for generating random graphs with planted community structure~\citep{holland1983stochastic}. In the SBM, a graph with $n$ vertices is partitioned into $K$ different communities. Each vertex is assigned to one of these $K$ communities, and the probability of an edge existing between any two vertices depends on their community memberships. Each vertex $v_i$ is labeled with a community assignment $y_i$ (where $y_i \in \{1,\ldots,K\}$). The probability of an edge forming between two vertices $v_i$ and $v_j$ is determined by their community assignments $y_i$ and $y_j$, according to a $K \times K$ probability matrix $\mathbf{P}$. This matrix $\mathbf{P}$ is symmetric and specifies the connection probabilities both within and between communities. To create graphs where vertices are more likely to connect with others in their same community, the diagonal elements of $\mathbf{P}$ (representing within-community connections) are set higher than the off-diagonal elements (representing between-community connections). This property has made the SBM a valuable tool for both theoretical analysis and empirical evaluation of graph clustering algorithms. However, compared to LFR benchmark graphs, homogeneity of SBMs isn't representative of the real world graphs. To analyze how GNNs perform community detection on SBMs, we utilize the Attributed Degree-Corrected Stochastic Block Model (ADC-SBM)~\citep{tsitsulin2022synthetic}. This framework generates synthetic graphs with, community structure, and node attributes. The graph generation process begins with $n$ nodes that are partitioned into $K$ communities, where community assignments are drawn from a uniform random distribution.  For notational clarity, we re-parameterize $d_{out}$ in Tsitsulin et al.~\citep{tsitsulin2022synthetic}, which controls the community complexity, as $\rho$. $\rho$ controls the expected number of inter-community connections per node in the ADC-SBM framework. When it is small, nodes primarily connect to others within their own community, creating strong, well-separated community structures that are easy to detect. As $\rho$ increases, nodes form more connections across community boundaries while maintaining fewer internal connections, progressively weakening the community structure. For experiments on ADC-SBM we explore network sizes of $10^3$ nodes with 10 clusters, and use the same parameters in Tstisulin et al.~\cite{tsitsulin2023graph}. We report the results for ADC-SBM in Appendix C. 


\subsection{\label{sec:level2}Synthetic node attributes}

To generate LFR graphs for characterizing the robustness of GNNs for community detection, we add attributes that segment the graphs into non-overlapping clusters. The attribute generation process begins by creating $K$ zero-mean attribute community centers, drawn from a $d$-dimensional multivariate normal distribution with covariance matrix $\Sigma_{K}= \sigma_c^2 I_n$. For each node belonging to $k^{th}$ community its attribute is generated from a $d$-dimensional multivariate normal distribution with covariance $\Sigma = \sigma^2 I_n $. The ratio $\frac{\sigma_c^2}{\sigma^2}$ determines the intra-community gap in the attribute space, where small ratio implies more complex community structures. In our configuration of the attributed LFR model, for the number of communities generated by LFR we set $d = 32$ dimensional attributes. We vary $\sigma_c \in (1,20)$ for the intra-cluster center variance and maintain $\sigma = 2$ for the within-cluster variance. For a comprehensive evaluation we  broaden the scope of the experimental design detailed in Tsitsulin et al.~\citep{tsitsulin2023graph}. 

\subsection{Real world networks}

\begin{table*}[!hbt]
     \caption{Summary of real world networks.\label{tab:datasets}}
     \addtolength{\tabcolsep}{12pt}   
     \centering
    \begin{tabular}{c c c c c }
    \toprule
        Dataset & Number of Nodes & Number of Edges & Attributes & Communities \\ \hline
        Cora & 2,708 & 10,556 & 1,433 & 7   \\ 
        Citeseer & 3,327 & 9,104 & 3,703 & 6  \\ 
        Pubmed & 19,717 & 88,648 & 500 & 3   \\ 
        \bottomrule
    \end{tabular}
\end{table*}
In addition to synthetic graphs, we also use real world citation networks~\citep{planetoid}. 
Specifically, we use Cora, Citeseer, and Pubmed datasets. Each network is an attributed graph where each node attribute is a bag-of-words vector representation of a corpus of scientific documents and citation links between them. 
The adjacency matrix $\mathbf{A}[i,j] = \mathbf{A}[j,i] = 1$ for $(i,j)$ pair of nodes when $i$ cites $j$.
We summarize details of the datasets used in Table~\ref{tab:datasets}.

\subsection{\label{sec:level2}Attacks on graphs}

We design an evaluation routine that tests robustness of GNN architectures for community detection under various attacks focusing on three categories of perturbations to the attributes. We evaluate community robustness under (1) node attribute perturbations, (2) graph topology perturbations, and (3) more sophisticated network attacks that target the learning process of the models (i.e. Nettack~\citep{nettack} and Metattack~\citep{mettack}). 

\subsubsection{Node attribute perturbation} Here, we perturb node attributes by first, altering the location and then scale of their distribution. We add random Gaussian noise generated by varying location parameter and keeping scale parameter fixed and then varying the scale parameter, but keeping the location parameter fixed, similar to ~\citep{tsitsulin2023graph}. By doing this, we perturb the distribution of attributes within each community potentially creating overlaps between clusters distributions during testing~\cite{nodeattacks}. For our naive approach, let $\mathbf{X} \in \mathbb{R}^{n\times d}$ be the node attribute attribute matrix for a graph $\mathcal{G}$. Then we generate an adversarial attribute matrix $\mathbf{X}_{adv} = \mathbf{X} + \varepsilon$ where, $\varepsilon$ is a matrix $\in \mathbb{R}^{n\times d}$ with each row as a multivariate-normal noise with mean $\mathbf{0}\in \mathbb{R}^d$ and variance $\Sigma = \sigma^2 I_d$.

\subsubsection{Graph topology perturbation}We explore how removing edges affects network structure through two approaches,  random and targeted edge deletions. We follow the procedure outlined by~\cite{wei2024robustness}. In the random approach, which emulates real-world network failures, we randomly select some percentage of nodes ranging between 10\% to 70\% of the nodes in increments of 10\% and remove their connected edges. Specifically, we create 50 independent realizations of the LFR network for each setting. In all realizations, we ensure network connectivity after edge removal procedure, which is essential for community detection. If disconnection occurs after removing edges, we discard that network and generate a new LFR and perform the edge removal procedure again. 

For targeted edge removal, we utilize node betweenness centrality, which quantifies a node's importance based on its frequency in shortest paths with other nodes. Nodes with high betweenness centrality are crucial network components, and removing their adjacent edges can significantly impact network functionality. This method simulates real-world adversarial attack. Implementation involves ranking nodes by betweenness centrality (1 for lowest, N for highest, where N is total nodes). We generate 50 independent node sets, each containing a proportion of the nodes, selected probabilistically based on betweenness centrality rank. We maintain network connectivity and recalculate community structure similarity ECS and take an averagescores across all sets.


\subsubsection{Adversarial attacks for community detection}
We also evaluate GNN robustness under more sophisticated attacks that target the overall learning process to degrade community detection performance. Specifically, we use Nettack~\citep{nettack} and Metattack~\citep{mettack} schemes, which represent advanced adversarial techniques designed to compromise graph-based machine learning systems while maintaining the appearance of legitimate network data.
Nettack operates by strategically perturbing both graph topology and node attributes in a manner that renders the attack virtually undetectable due to the deliberately small magnitude of its modifications. The attack methodology employs a surrogate model to systematically measure the impact of potential modifications, subsequently applying those perturbations that yield maximum impact on specifically targeted nodes. The fundamental objective involves modifying the original graph to create a perturbed version while constraining the total number of changes through a predefined budget that limits alterations to both node attributes and edge connections across the entire network.
To ensure that perturbations remain unnoticeable to detection mechanisms, Nettack incorporates two critical preservation strategies. For structural modifications, the attack maintains the power-law degree distribution characteristic of real-world networks through rigorous likelihood ratio testing, ensuring that the modified graph retains statistical properties consistent with naturally occurring networks. This preservation mechanism prevents the attack from creating obviously artificial connectivity patterns that would immediately flag the network as compromised. For attribute perturbations, Nettack implements a sophisticated co-occurrence preservation mechanism based on random walks across the attribute co-occurrence graph, ensuring that newly added attributes maintain realistic relationships with existing node attributes. This mechanism restricts attribute additions to only those that demonstrate sufficient co-occurrence probability with the node's original attribute set, thereby maintaining the semantic coherence of node representations.
The attack framework is formulated as a bi-level optimization problem that utilizes a surrogate model based on GNNs. Such a surrogate model enables efficient computation of attack gradients while approximating the behavior of more complex target models. Empirical evaluations on tasks such as node classification and link prediction~\citep{nettack} demonstrate that Nettack successfully reduces prediction accuracy on node classification tasks without compromising the fundamental structural and statistical properties of graphs.\\
Metattack represents a gradient-based adversarial approach that perturbs data during the training phase with the explicit goal of reducing the effectiveness of embeddings learned by the target network. This attack is formulated as a bi-level optimization problem where the adversarial model first trains on the perturbed graph and subsequently maximizes the node community assignment loss over a validation set. This sophisticated approach creates an attack designed to minimize its detectability while maximizing its impact on the target system's performance. The meta-gradient approach addresses this challenge by treating the graph structure itself as a hyperparameter subject to optimization, enabling the attack to learn optimal perturbation strategies through the training process.
Metattack aims to preserve key graph properties through multiple constraint mechanisms to maintain attack stealth. The attack employs degree distribution preservation through likelihood ratio testing, ensuring that the modified network retains the scale-free properties typical of real-world graphs. Additionally, the attack operates under strict budget constraints that limit the total number of perturbations across both node attributes and edge modifications. This dual constraint system ensures that the attack remains undetectable while still achieving significant degradation in target model performance. The meta-gradient computation enables the attack to optimize perturbations by considering their effects throughout the entire training trajectory, making Metattack particularly effective against adaptive defense mechanisms that might detect simpler attack strategies.
\subsection{\label{sec:level2}Graph community detection with GNNs}

 GNNs perform nonlinear attribute aggregation based on the graph structure. They rely on two fundamental operations: message passing and pooling. Message passing enables information flow between nodes, where nodes aggregate information from their neighborhoods through differentiable functions to capture graph structure. Once node representations are learned through multiple message passing iterations, pooling operations can be used to aggregate these representations to obtain a graph-level embedding. This enables fixed-size embeddings for graphs of varying sizes and structures, facilitating tasks such as graph classification, regression, and similarity comparison, with common pooling approaches including global summation/averaging, hierarchical clustering, and attention-weighted aggregation~\cite{souravlas2021survey, sobolevsky2022graph,liu2022survey,leeney2024uncertainty,kumar2024analyzing}.

\begin{figure}
    \centering
    \includegraphics[width=0.48\textwidth,height = 13cm]{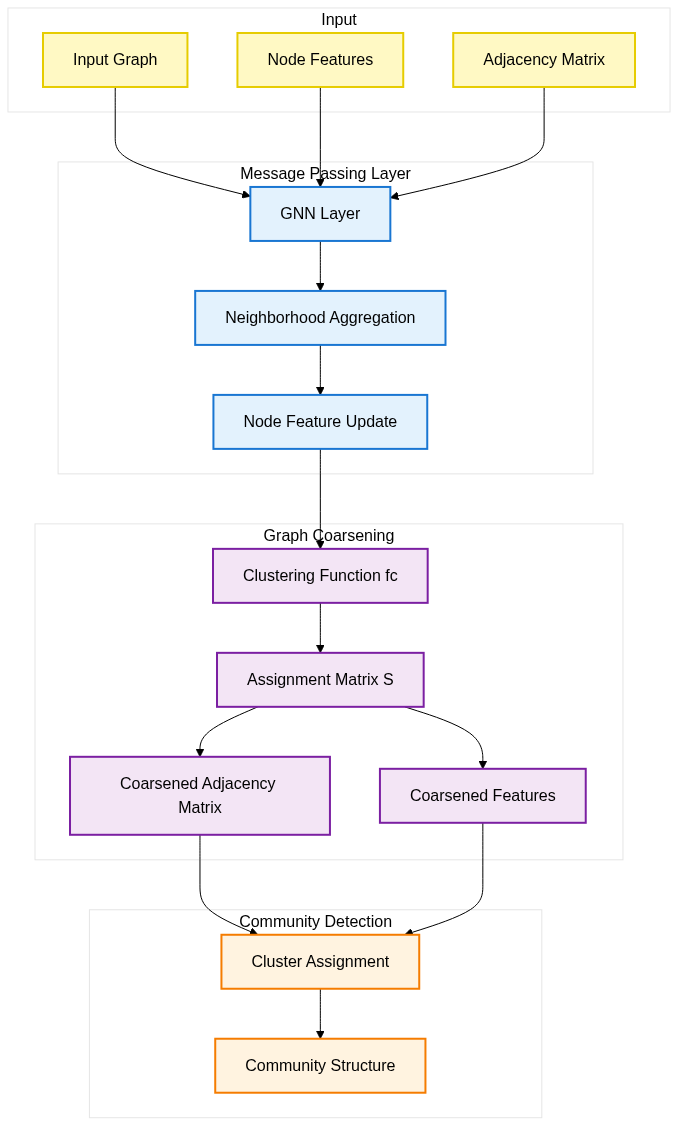}
    \caption{GNN based community detection approach takes node attributes and the graph adjacency as inputs. Based on quality functions and graph partition metrics, it iteratively learns graph embeddings for nodes that are assigned to communities.}
    \label{fig:architecture_diag}
\end{figure}

From Fig.~\ref{fig:architecture_diag} we can see the general pipeline on how community detection is performed for GNNs. GNNs are used to learn node representations based on their attributes as well as the adjacency and then clustering is performed on the learnt embeddings. The community assignment can be obtained via a softmax function applied to the GNN output:
\[
\mathbf{C} = \text{softmax}(\text{GNN}(\tilde{\mathbf{A}}, \mathbf{X})),
\]
where the general model for GNN first proposed by~\cite{merkwirth2005automatic,scarselli2008graph} can be given by 
\begin{equation*}
    h_u^{(i)} = \sigma\left(\mathbf{W}_{\text{self}}^{(i)}h_u^{(i-1)} + \mathbf{W}_{\text{neigh}}^{(i)}\sum_{v\in N(u)}h_v^{(i-1)} + b^{(i)}\right),
\end{equation*}
where $h_u^{(t)}$ is the embedding vector of node $u$ at layer $t$; $\mathbf{W}_{\text{self}}^{(t)} \in \mathbb{R}^{d^{(t)} \times d^{(t-1)}}$ and $\mathbf{W}_{\text{neigh}}^{(t)} \in \mathbb{R}^{d^{(t)} \times d^{(t-1)}}$ is are learnable weight matrices and $\sigma(\cdot)$ is some choice of non-linearity like ReLU or tanh. 
This allows for end-to-end learning of graph clusterings that leverage both structural and attribute information. The challenge lies in designing appropriate differentiable objective functions that can optimize for high-quality clusters while being amenable to gradient-based optimization in the GNN framework.

Graph coarsening approaches enhance the pooling process by incorporating graph topology, addressing key limitations of set pooling methods. The core concept revolves around a clustering function $f_c: \mathcal{G} \times \mathbb{R}^{|V| \times d} \rightarrow \mathbb{R}^{|V| \times c}$ that maps nodes to cluster assignments. This function generates an assignment matrix $\mathbf{S} = f_c(G)$, where $\mathbf{S}[u,i] \in \mathbb{R}$ quantifies the association between node $u$ and cluster $i$. The critical aspect of this approach lies in its differentiability requirement. The clustering function must be differentiable for end-to-end training, which excludes conventional clustering algorithms like spectral clustering. The coarsening process then transforms the graph through two fundamental operations: (1) computing a new coarsened adjacency matrix $\mathbf{A}_{new} = \mathbf{S}^\top \mathbf{A} \mathbf{S} \in \mathbb{R}_+^{c \times c}$ and (2) generating a new attribute matrix $\mathbf{X}_{new} = \mathbf{S}^\top \mathbf{X} \in \mathbb{R}^{c \times d}$. These transformations create a representation where $\mathbf{A}_{new}$ captures inter-cluster edge strengths and $\mathbf{X}_{new}$ contains aggregated node embeddings for each cluster. This process is applied iteratively, progressively reducing the graph size, with the final graph representation obtained through set pooling on the sufficiently coarsened graph. 

Quality functions and cluster distance metrics help determine boundaries by which suitable partitions should be made on graphs~\citep{fortunato2010community}. They measure how well-connected nodes are within communities versus between communities, offering guidance about community structure which is particularly important for GNNs. Such objectives enables iterative learning of GNN-based clustering algorithms.

GNNs for community detection can be categorized into two fundamental paradigms based on their learning approach and architectural design~\cite{liu2022survey,su2022comprehensive}. Supervised methods (e.g., GCN, GAT, GraphSAGE) leverage ground truth community labels during training to learn node representations through message passing mechanisms. In contrast, unsupervised methods (e.g., DiffPool, MinCUT, DMoN) perform end-to-end community detection without requiring labeled data by optimizing graph-theoretic objectives through differentiable pooling mechanisms. Specifically, DiffPool implements hierarchical graph coarsening using dual GNN architectures for embedding and cluster assignment. MinCUT adapts spectral clustering principles through differentiable normalized cut optimization. DMoN combines modularity maximization with collapse regularization to prevent degenerate solutions. The fundamental distinction lies in supervised methods requiring pre-defined community labels to learn classification boundaries, while unsupervised approaches discover community structure through intrinsic graph properties like hierarchical pooling, spectral characteristics, or modularity, making them applicable when ground truth communities are unknown or unavailable. Details on the quality functions used to optimize unsupervised objective of recovering community structure can be found in Appendix A.

\subsection{\label{sec:level2}Community similarity} We use element centric similarity metric~\citep{gates2019element} for community similarity. ECS handles disjoint, overlapping, or hierarchically structured communities. It operates by examining how nodes are grouped across different communities, utilizing a PageRank process to capture relationships between nodes within communities. The similarity between clusters is then calculated by comparing the PageRank affinity for each element. ECS is a versatile and an interpretable metric that can illuminate community comparison at the node level, without requiring computationally inefficient  community matching. ECS produces a score between 0 and 1 where the closer it is to one, the more similar two communities are. This measure of community similarity overcomes traditional biases present in other similarity measures while providing a unified framework for comparing disjoint, overlapping, and hierarchically structured communities. More details on ECS are provided in Appendix B.

\section{\label{sec:Results}Results and Discussion}

We present a comprehensive evaluation of community detection robustness across six GNN architectures in two distinct categories. First, in Sec.~\ref{sec:sup}, we report results on supervised message-passing mechanisms (i.e. GCN, GAT and GraphSAGE), which serve as the most widely used building blocks in GNNs. Second, in Sec.~\ref{sec:unsup}, we analyze specialized clustering approaches that employ pooling mechanisms, namely DiffPool, MinCut, and DMoN, which are specifically designed to identify community structures in a completely unsupervised manner. We summarize the results on node and topology perturbations in Sec.~\ref{sec:emp_summary}.

In Sec.~\ref{sec:adversarial}, we then analyse modern adversarial attacks (i.e., Nettack and Metattack) which are understudied for community detection performance. In Sec.~\ref{sec:citation_net}, we extend our analysis to real-world citation networks. Our comprehensive evaluation illuminates a fundamental trade-off. Supervised methods excel in ideal conditions but deteriorate rapidly when faced with attribute noise or structural attacks, whereas unsupervised approaches, particularly, have more consistent performance across varying perturbation intensities. 

In the subsequent figures (Figs. 2-10), unless otherwise specified, each set contains four subfigures arranged in a 2×2 grid presenting different perturbation scenarios. The top row, i.e., subfigures (a) and (b), show results for node attribute perturbations, where (a) displays the impact of location perturbations and (b) illustrates scale perturbations. Similarly, the bottom row, i.e., subfigures (c) and (d), present structural perturbation results, with (c) showing random edge deletions and (d) depicting targeted edge removals based on betweenness centrality. In all figures, the x-axis represents the perturbation intensity, while the y-axis shows the mean ECS (taken over 50 runs). We do not report standard deviation here as they are negligible with respect to the mean. Table~\ref{tab:pertub_setup} summarizes the details of the perturbation experiments. We also provide our code at \cite{github_code} and hyper-parameter details for reproducibility of our results in Appendix D.

\begin{table}[!hbt]
\caption{Three categories of perturbations and their parameters.}
\label{tab:pertub_setup}
\centering
\begin{tabular}{lll}
\toprule
\textbf{Category} & \textbf{Experiment Type} & \textbf{Parameters} \\
\midrule
\multirow{2}{*}{Node Attribute} & Location & 0, 1, 2, 3, 4 \\
& Scale& 1, 5, 10, 15, 20 \\
\midrule
\multirow{2}{*}{Edge Deletions} & Random & 1\% to 70\% \\
& Targeted & 1\% to 70\% \\
\midrule
\multirow{2}{*}{Adversarial Attacks} & Nettack & Default settings \\
& Metattack & Default settings \\
\bottomrule
\end{tabular}
\end{table}

\begin{table*}[!ht]
    \centering
    \addtolength{\tabcolsep}{10pt} 
    \begin{tabular}{ll@{\hspace{3cm}}p{5cm}}
        \hline
        \textbf{Parameter} & \textbf{Description} & \textbf{Value} \\
        \hline
        $N$ & number of nodes & 1,000 \\
        $maxk$ & max degree for LFR & $0.1N$ \\
        $\langle k \rangle$ & average degree for LFR & 25 \\
        $maxc$ & max community size for LFR & $0.1N$ \\
        $\alpha$ & degree distribution exponent for LFR & $-2$ \\
        $\beta$ & community size distribution exponent for LFR & $-1.1$ \\
        $\mu$ & mixing parameter for LFR &  0.1, 0.2, 0.3, 0.4, 0.5 \\
        $r$ & number of realizations & 50 \\
        \hline
    \end{tabular}
    \caption{LFR benchmark parameters and their values.}
    \label{tab:lfr-params}
\end{table*}

\subsection{\label{sec:sup}Supervised community detection}

\subsubsection{\label{sec:GCN}GCN}
GCNs are a type of GNN designed to work with graph structured data, extending the concept of convolutional neural networks to graphs. They allow the learning of node representations based on both node features and graph structure. The core operation in a GCN is the graph convolution, where for a single GCN layer, the output features $H^{(l+1)}$ are computed using the formula $H^{(l+1)} = \sigma(\mathbf{D}^{-\frac{1}{2}}\mathbf{A}\mathbf{D}^{-\frac{1}{2}}H^{(l)}\mathbf{W}^{(l)})$. Here, $H^{(l)}$ is the input feature matrix from the previous layer, $\mathbf{A}$ is the adjacency matrix, $\mathbf{D}$ is the degree matrix of $\mathbf{A}$, $\mathbf{W}^{(l)}$ is the trainable weight matrix, and $\sigma$ is a non-linear activation function. The term $\mathbf{D}^{-\frac{1}{2}}\mathbf{A}\mathbf{D}^{-\frac{1}{2}}$ is the normalized adjacency matrix, which helps prevent numerical instabilities and exploding/vanishing gradients. 

GCNs are typically trained using gradient descent to minimize a task specific loss function, such as cross-entropy for node classification tasks. This formulation allows GCNs to learn representations that capture both local and global graph structure, making them powerful tools for various graph-based machine learning tasks.
\begin{figure}[h]
    \centering
    \includegraphics[width=0.48\textwidth]{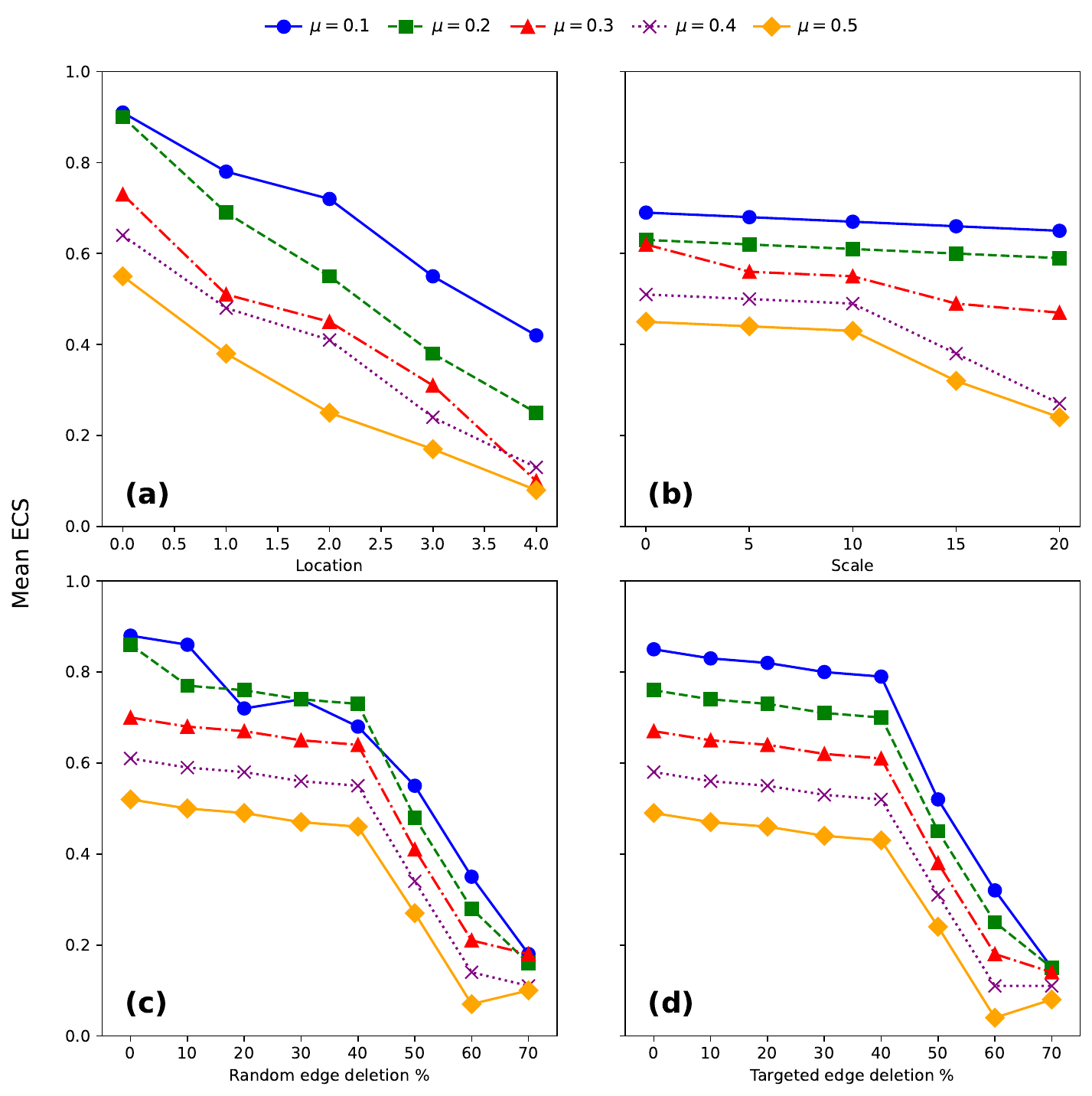}
    \caption{ECS scores for GCN-based community detection under four perturbation scenarios. Under node attribute perturbations, (a) shows ECS score across different means and (b) represents scale perturbations. Under structural perturbations, (c) shows results for random edge perturbations and (d) for targeted attacks based on betweenness centrality.}
    \label{fig:GCN}
\end{figure}

Our experimental results demonstrate significant variations in how GNN-based community detection methods respond to different types of perturbations across networks with varying community strengths.
Fig~\ref{fig:GCN} (a) shows that performance degrades as location perturbations to node attributes increases, with this effect being strongly modulated by the underlying community strength. Networks with stronger community structure ($\mu = 0.1$, $\mu = 0.2 $) maintain ECS scores above 0.4 even at the highest perturbation levels, while networks with weaker communities ($\mu = 0.4$, $\mu = 0.5$) deteriorate rapidly, falling below 0.2. This suggests that strong community partitioning may partially compensate for noisy node features.
The scale analysis in Fig.  2(b) further supports this finding, showing relatively stable performance for stronger community structures across varying degrees of scale perturbation levels. In contrast, networks with $\mu = 0.4$ and $\mu = 0.5$ weak ECS, showing complex community structures can get impacted by higher levels of perturbation to attributes scale parameter. 

For random edge deletions Fig.  2(c), shows two things, initial robustness followed by rapid collapse. Performance remains relatively stable until 30-50\% 
edge deletions, after which there is significant drops across all $\mu$. Interestingly, for network with mixing parameter $\mu = 0.2$ occasionally outperforms $\mu = 0.1$ in the mid-range of deletions, suggesting that moderate community strength might provide some robustness to GCN. 
Although targeted edge deletions based on betweenness centrality show a more gradual initial decline, community recovery drops from 0.88 to 0.18 ECS, a 79\%  drop for targetted edge deletions. We observe that even for graphs with strongest community structures ($\mu = 0.1$), they cannot be effectively recovered at higher levels of attack with performance dropping from 0.88 to 0.18 as significant edges are deleted.

\subsubsection{\label{sec:gat}GAT}
GAT uses attention mechanism enabling structural learning on networks. In this GNN, the node features are updated as $h_i^{(l+1)} = \sigma(\sum_{j \in \mathcal{N}(i)} \alpha_{ij}\mathbf{W}^{(l)}h_j^{(l)})$, with attention coefficients $\alpha_{ij} = \text{softmax}_j(e_{ij}) = \frac{\exp(e_{ij})}{\sum_{k \in \mathcal{N}(i)} \exp(e_{ik})}$, where $e_{ij} = \text{LeakyReLU}(\mathbf{a}^T[\mathbf{W}^{(l)}H_i^{(l)} \| \mathbf{W}^{(l)}H_j^{(l)}])$ and $\|$ denotes concatenation. This allows the network to assign different importance to different neighbors when aggregating information.

\begin{figure}[h]
    \centering
    \includegraphics[width=0.48\textwidth]{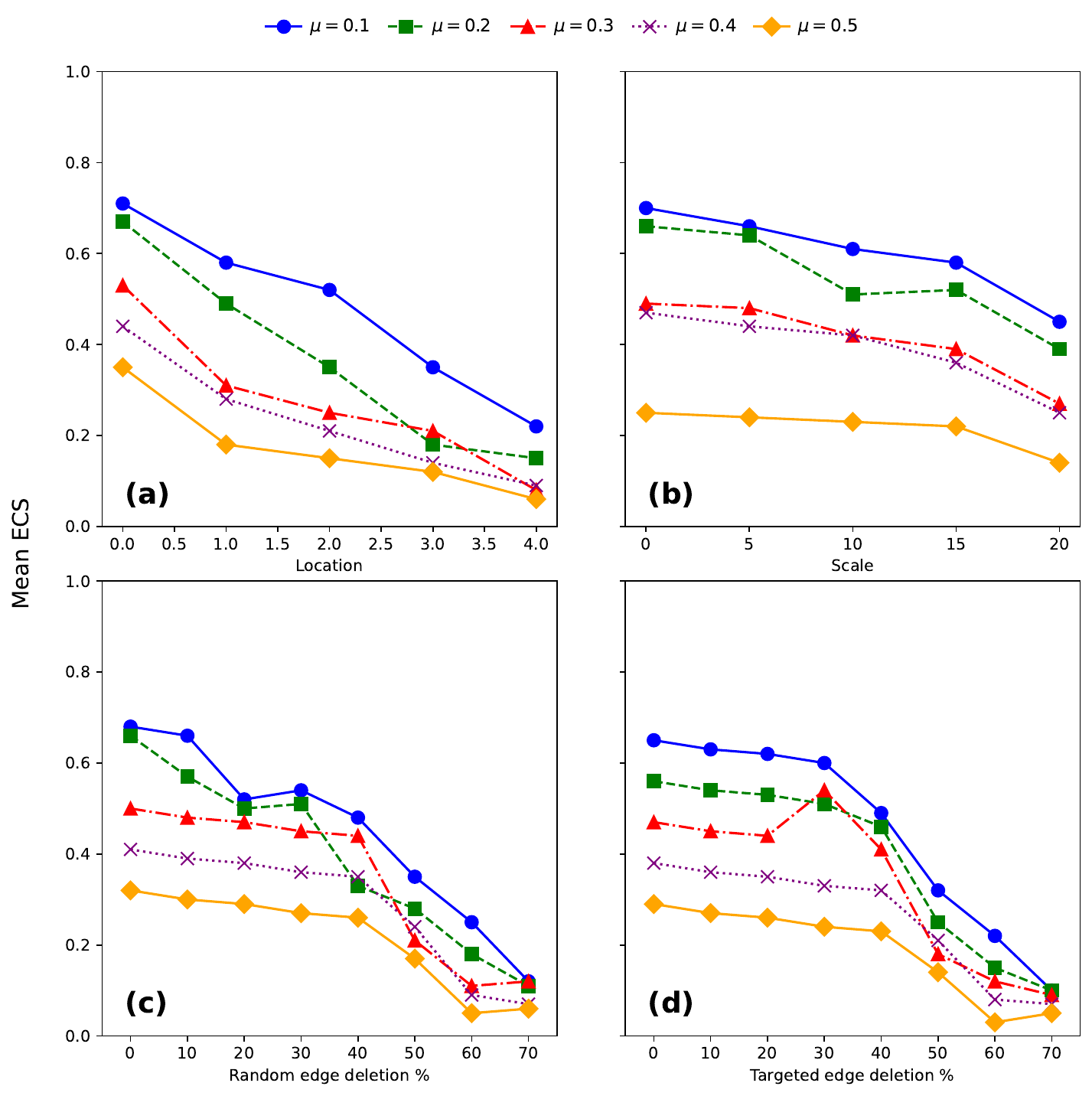}
    \caption{ECS scores for GAT-based community detection under four perturbation scenarios. Under node attribute perturbations, (a) shows ECS score across different means and (b) represents scale perturbations. Under structural perturbations, (c) shows results for random edge perturbations and (d) for targeted attacks based on betweenness centrality.}
    \label{fig:GAT}
\end{figure}

GAT demonstrates relatively poor robustness across all perturbation types. For attribute perturbations, location shifts to attribute distribution cause large degradation with a 78\% average drop in community recovery performance. The decline is particularly larger for weaker communities where drops range from 69\% ($\mu$=0.1) to 84.9\% ($\mu$=0.5), with higher drops at even low perturbation levels indicating rapid failure onset. Scale shifts on the other hand show considerably better robustness with a 42.5\% average drop in community recovery across different levels of $\mu$, maintaining more consistent performance with drops ranging only from 35.7\% to 46.8\% and exhibiting gradual degradation patterns. Graph topology perturbations through edge deletions lead to a large performance degradation for both random (ranging 76-83.3\%) and targeted (ranging 80.9-84.6\%) deletions, with targeted attacks showing slightly worse but consistently high vulnerability across all community strength levels. This suggests that GATs community recovery capability tends to rely more on structural integrity.

GAT's baseline performance strongly correlates with community strength parameters, declining from 0.685 average ECS for the strongest communities ($\mu$ = 0.1) to just 0.302 for the weakest ($\mu$ = 0.5), indicating that the model struggles even with clean data when community boundaries become less distinct. The poor robustness of GAT might be explained by the fact that its attention mechanism heavily relies on both graph topology and node feature distributions. Location shifts alter the feature space geometry that attention weights depend on, fundamentally disrupting the learned attention patterns and causing cascading failures in community boundary detection, while edge deletions directly disrupt computation of robust attention weights. This implies that the model appears to overfit to specific training distributions. Variance shifts show better robustness because they preserve relative feature relationships and distributional structure that attention mechanisms can still leverage, maintaining the geometric properties of the feature space that GAT's attention computation relies upon.

\subsubsection{\label{sec:sage}GraphSAGE}
GraphSAGE generalizes the neighborhood aggregation approach with $h_i^{(l+1)} = \sigma({\bf W}^{(l)} \cdot \text{CONCAT}(h_i^{(l)}, \text{AGG}(\{h_j^{(l)}, \forall j \in N(i)\})))$, where AGG is a permutation-invariant aggregation function such as mean, max, or LSTM, applied to the neighbor set. The key innovation is the sampling of a fixed-size neighborhood and the combination of the node's own features with the aggregated neighborhood features.

\begin{figure}[h]
    \centering
    \includegraphics[width=0.48\textwidth]{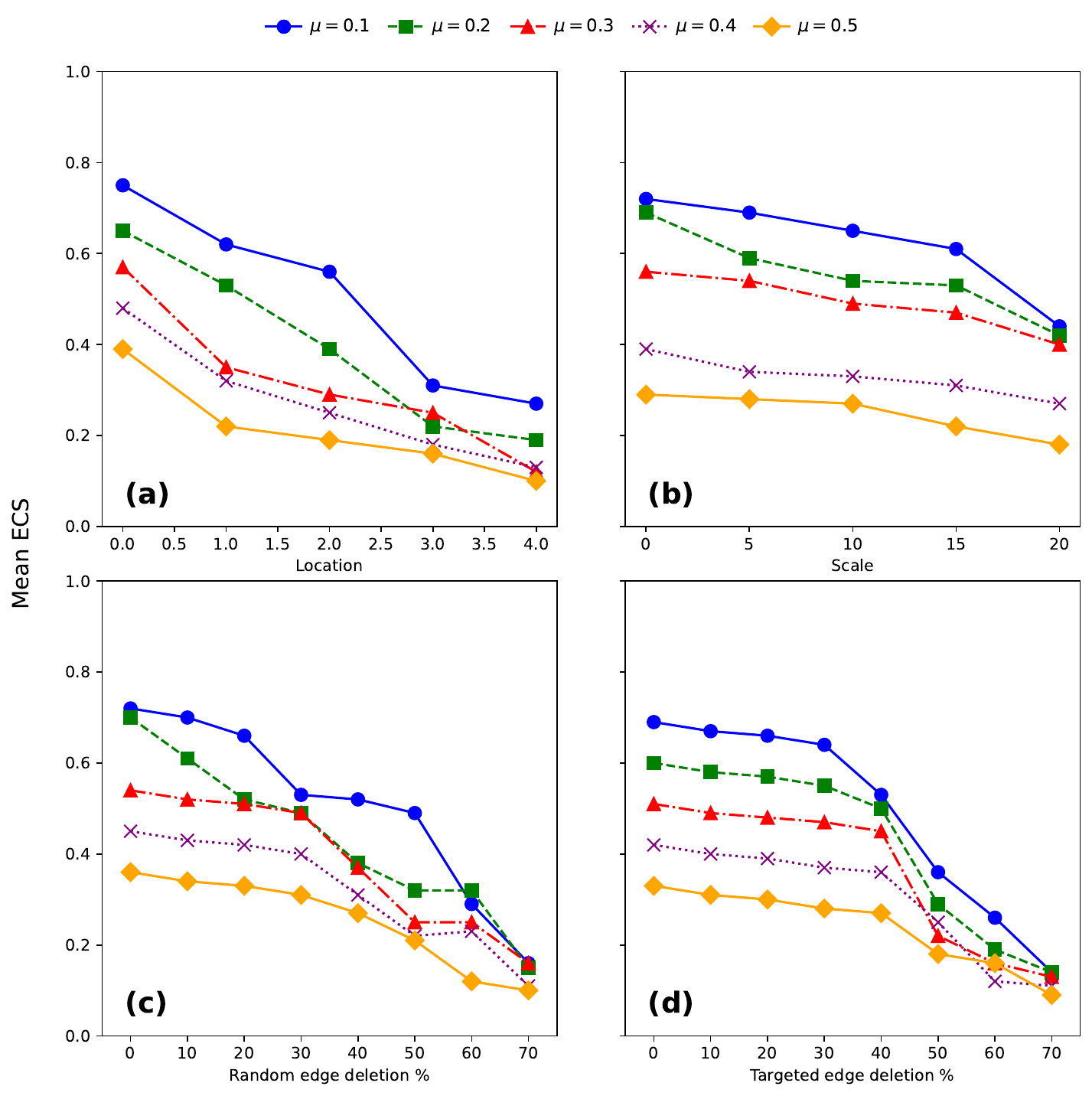}
    \caption{ECS scores for SAGE-based community detection under four perturbation scenarios. Under node attribute perturbations, (a) shows ECS score across different means and (b) represents scale perturbations. Under structural perturbations, (c) shows results for random edge perturbations and (d) for targeted attacks based on betweenness centrality.}
    \label{fig:SAGE}
\end{figure}

GraphSAGE demonstrates moderate robustness relative to GAT. For attribute perturbations, location shifts exhibit declines that range from 64\% ($\mu$=0.1) to 78.9\% ($\mu$=0.3), with initial drops ranging from 17-44\% indicating that GraphSAGE can be sensitive to location shifts at even small perturbations. Scale shifts show significantly better robustness across all community strengths maintaining relatively consistent performance with drops ranging from 28.6\% to 39.1\% and exhibiting small losses of just 3-15\% for small levels of perturbations. This may suggest that GraphSAGEs sampling-based aggregation provides better protection against variance changes. Graph topology perturbations through edge deletions lead to substantial performance degradation for both random (ranging 70.4-78.6\%) and targeted (ranging 72.7-79.7\%) deletions, with targeted attacks showing marginally worse performance, but relatively consistent degradation across all community strength levels. Notably, though for both types of topology perturbations, GraphSAGE shows smaller drops at lower levels of edge deletions (2-13\%) indicating more gradual degradation patterns than attribute perturbations, making it more robust for edge attacks. 

GraphSAGE maintains a stronger baseline performance at different levels of $\mu$, gradually declining from an average ECS of 0.72 for the strongest community($\mu$ = 0.1) to 0.34 for the weakest ($\mu$ = 0.5). This suggests better handling of ambiguous community boundaries compared to GAT. The superior robustness of GraphSAGE maybe attributed to its sampling-based aggregation mechanism that provides inherent regularization against perturbation that is, the random neighbor sampling creates natural noise resistance, while the hierarchical aggregation across multiple hops maintains structural information even when local topology is disrupted. Unlike attention-based mechanisms that can be severely disrupted by feature distribution changes, GraphSAGE's aggregation functions are more robust to location shifts in the feature space. The sampling approach also provides some resilience to edge deletions since the model does not rely on exact neighborhood preservation, though performance still degrades substantially when critical structural information is lost through extensive edge removal.

\subsection{\label{sec:unsup}Unsupervised clustering}
\subsubsection{\label{sec:MinCut}MinCut}

MinCutPool is differentiable graph pooling layer that performs hierarchical clustering based on spectral clustering~\citep{bianchi2020spectral_mincut}. The key idea is to formulate a continuous relaxation of the normalized MinCut problem and training a GNN to compute optimal cluster assignments.
The normalized MinCut objective aims to partition a graph's vertices $V$ into $K$ disjoint subsets by minimizing number of edges removed. This can be expressed as maximizing:
\[\frac{1}{K}\sum_{k=1}^K \frac{\text{links}(V_k)}{\text{degree}(V_k)} = \frac{1}{K}\sum_{k=1}^K \frac{\sum_{i,j\in V_k} E_{i,j}}{\sum_{i\in V_k,j\in V\setminus V_k} E_{i,j}}.\]
MinCutPool implements this through a cluster assignment matrix $\mathbf{S} \in \mathbb{R}^{N\times K}$ computed by applying message passing layers followed by an MLP with softmax activation.
The pooling operation produces a coarsened adjacency matrix and pooled vertex features. The model is trained by minimizing an unsupervised
orthogonality loss encouraging the cluster assignments to be orthogonal.
The key advantages of MinCutPool include end-to-end differentiability without requiring eigendecomposition, enabling learning embeddings for both graph structure and node features.

\begin{figure}[h]
    \centering
    \includegraphics[width=0.48\textwidth]{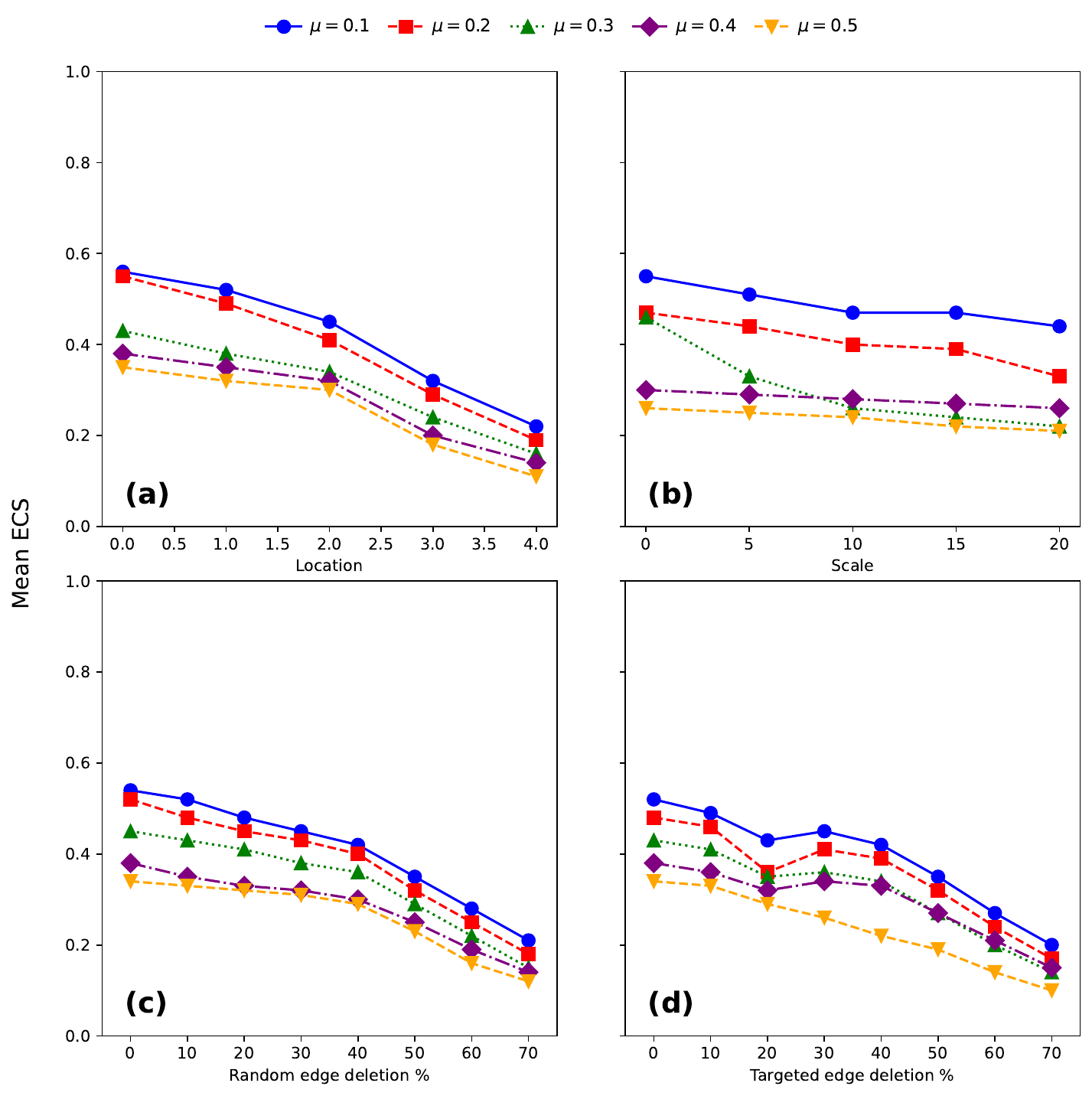}
    \caption{ECS scores for MinCut-based community detection under four perturbation scenarios. Under node attribute perturbations, (a) shows ECS score across different means and (b) represents scale perturbations. Under structural perturbations, (c) shows results for random edge perturbations and (d) for targeted attacks based on betweenness centrality.}
    \label{fig:MinCut}
\end{figure}

Fig.~\ref{fig:MinCut} shows varying levels of robustness across different perturbation scenarios for MinCut. In Fig. 5(a), location perturbation experiments on node attributes shows MinCut has strongest robustness at low mixing parameter ($\mu = 0.1$), achieving ECS of 0.56 which gradually deteriorates to 0.22 as perturbation increases. This significant drop (approximately 61\%) indicates that MinCut's performance is notably sensitive to attribute perturbations when location of attribute distribution is perturbed at greater degree. The scale analysis in Fig. 5(b) reveals an interesting pattern: at $\mu = 0.1$, the model maintains relatively stable performance (0.55 to 0.44), suggesting consistent behavior under mild perturbations. However, for higher mixing parameters ($\mu = 0.4$ and $\mu = 0.5$), while the absolute performance is lower, the scale remains more stable (ranging from 0.30 to 0.26 for $\mu = 0.4$), indicating more predictable, albeit reduced, performance under severe perturbations.
For edge manipulation scenarios, MinCut shows similar patterns to both random and targeted edge deletions, with better robustness against random deletions. Fig. 5(c) shows that for random edge deletions, ECS at $\mu = 0.1$ starts at 0.54 and gradually decreases to 0.21, while targeted deletions show a slightly steeper degradation from 0.52 to 0.20. This marginal difference suggests that MinCut's community detection capabilities are somewhat more robust against random structural changes compared to targeted attacks. The model performance consistently degrades as $\mu$ increases across all test scenarios, with the most severe impact observed in location perturbations and targeted edge deletions. This suggests that MinCut's community detection ability is particularly vulnerable to both attribute noise and structural attacks when the underlying community structure becomes less distinct (higher $\mu$ values in the range of 0.4-0.5).

\subsubsection{\label{sec:DiffPool}DiffPool}
DiffPool introduces hierarchical graph representation learning through differentiable pooling layers that enable end-to-end training of deep GNNs~\citep{ying2018hierarchical_diffpool}. Given a graph with adjacency matrix $\mathbf{A} \in [{0,1}]^{n\times n}$ and node features $\mathbf{X} \in \mathbb{R}^{n\times p}$. DiffPool learns to cluster nodes at each layer l using two GNNs
An embedding GNN generates node representations:
$$Z^{(l)} = \text{GNN}_{l,\text{embed}}(\mathbf{A}^{(l)}, \mathbf{X}^{(l)}),$$
a pooling GNN produces soft cluster assignments:
$$\mathbf{S}^{(l)} = \text{softmax}(\text{GNN}_{l,\text{pool}}(\mathbf{A}^{(l)}, \mathbf{X}^{(l)})).$$
The pooling operation coarsens the graph via:
$$\mathbf{X}^{(l+1)} = {\mathbf{S}^{(l)}}^T \mathbf{Z}^{(l)}$$
$$\mathbf{A}^{(l+1)} = {\mathbf{S}^{(l)}}^T \mathbf{A}^{(l)}\mathbf{S}^{(l)}.$$
The model is trained end-to-end with two auxiliary objectives. A link prediction loss enforces cluster locality:
$$L_{LP} = |\mathbf{A}^{(l)}, \mathbf{S}^{(l)}{\mathbf{S}^{(l)}}^T|_F,$$
and an entropy regularization encourages discrete assignments:
$$L_E = \frac{1}{N}\sum_{i=1}^N H(S_i).$$ 

\begin{figure}[h]
    \centering
    \includegraphics[width=0.48\textwidth]{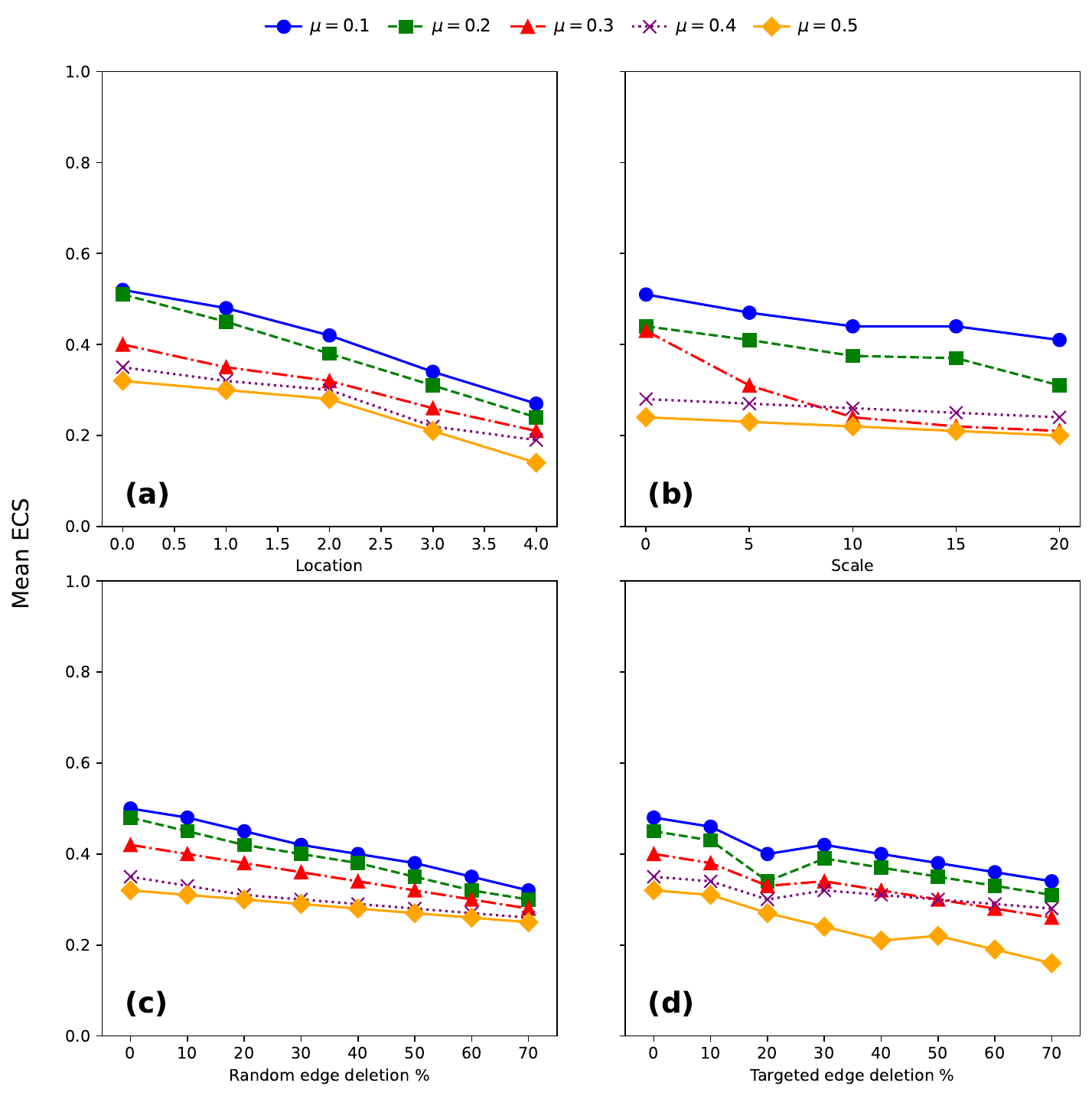}
    \caption{ECS scores for DiffPool-based community detection under four perturbation scenarios. Under node attribute perturbations, (a) shows ECS score across different location shifts and (b) represents scale perturbations. Under structural perturbations, (c) shows results for random edge perturbations and (d) for targeted attacks based on betweenness centrality.}
    \label{fig:DiffPool}
\end{figure}
Fig.~\ref{fig:DiffPool} shows that DiffPool exhibits decreasing robustness as the mixing parameter increases across all perturbation scenarios. For location perturbations experiments in Fig. 6(a), the model demonstrates its best performance at $\mu = 0.1$ with an ECS of 0.52, which progressively declines to 0.27 under increased perturbations. This substantial decrease (approximately 48\%) indicates DiffPool's sensitivity to attribute perturbations, though the degradation is less severe compared to MinCut.
The scale perturbation analysis in Fig.  6(b) shows that for $\mu = 0.1$, the model maintains relatively consistent performance (0.51 to 0.41), suggesting robust behavior under mild perturbations. For higher mixing parameters ($\mu = 0.4$ and $\mu = 0.5$), the scale stabilizes at lower values (around 0.24-0.28 for $\mu = 0.4$), indicating more predictable but reduced performance under severe perturbations.
In edge manipulation scenarios in Figs. 6(a) and 6(b), DiffPool shows interesting robustness patterns. For random edge deletions, the performance at $\mu = 0.1$ starts at 0.50 and shows a more gradual decline to 0.32, demonstrating better stability compared to mean performance. Similarly, targeted edge deletions show initial performance of 0.48 declining to 0.34, suggesting that DiffPool's hierarchical pooling structure provides some inherent protection against structural attacks.
A notable observation is that DiffPool maintains more consistent performance under random edge deletions compared to targeted attacks, particularly at higher $\mu$ values. This is especially evident at $\mu = 0.5$, where random edge deletion performance stays relatively stable (0.32 to 0.25), while targeted deletions show more dramatic degradation (0.32 to 0.16), indicating that the hierarchical structure of DiffPool is more vulnerable to targeted structural modifications than random perturbations.
Overall, while DiffPool shows lack of robustness to increasing $\mu$ values, it demonstrates better stability in edge deletion scenarios compared to its performance under attribute perturbations, suggesting that its hierarchical pooling mechanism provides some structural robustness despite being sensitive to node attribute noise.

\subsubsection{\label{sec:dmon}DMoN}

DMoN introduces an unsupervised clustering approach that leverages a differentiable modularity objective for graph pooling~\citep{tsitsulin2023graph}. The key idea is to encode community assignments using GNNs and a modularity based objective function to optimize to make a learnable graph clustering function in an end to end unsupervised manner. Using a softmax function for differentiabibility, soft community assignments, 
$\mathbf{C} = \text{softmax}(\text{GNN}(\mathbf{A}, \mathbf{X}))$ are then learnt using node attributes and graph topology. 
The clustering objective combines modularity maximization with a collapse regularizer:
$$\mathcal{L}_{\text{DMoN}}(\mathbf{C};\mathbf{A}) = -\frac{1}{2m}\text{Tr}(\mathbf{C}^T \mathbf{B}\mathbf{C}) + \frac{\sqrt{k}}{N}\left|\sum_i \mathbf{C}_i^T\right|_F - 1$$
where $\mathbf{B}$ is the modularity matrix:
$$\mathbf{B} = \mathbf{A} - \frac{\mathbf{d}\mathbf{d}^T}{2m},$$
and $\mathbf{d}$ being the degree vector.
The first term maximizes modularity by encouraging densely connected clusters. The second term prevents degenerate solutions by penalizing unbalanced cluster sizes.

\begin{figure}[h]
    \centering
    \includegraphics[width=0.48\textwidth]{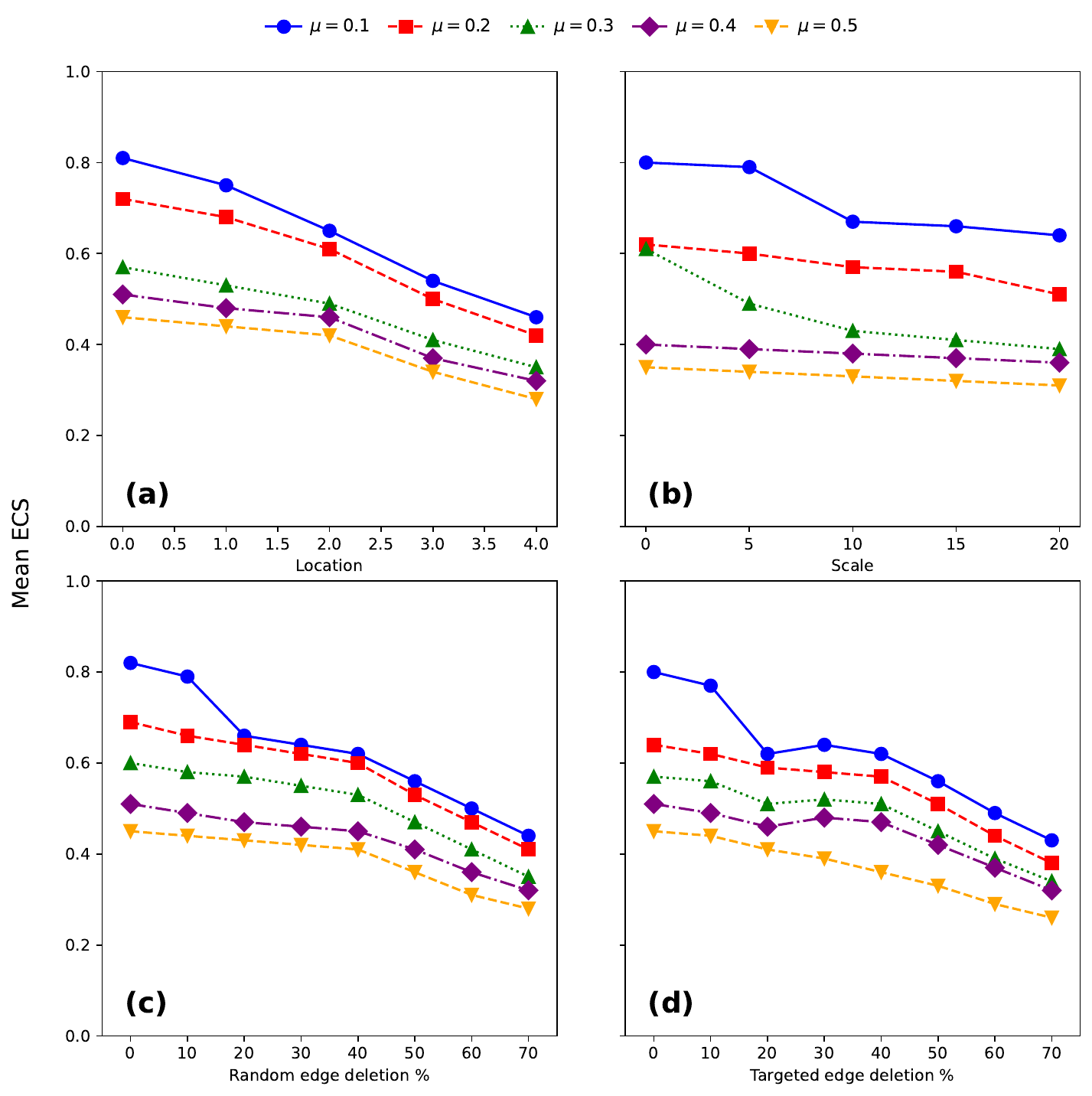}
    \caption{ECS scores for DMoN-based community detection under four perturbation scenarios. Under node attribute perturbations, (a) shows ECS score across different means and (b) represents scale perturbations. Under structural perturbations, (c) shows results for random edge perturbations and (d) for targeted attacks based on betweenness centrality.}
    \label{fig:DMoN}
\end{figure}
Robustness of DMoN architecture for community detection when location perturbations are done to node attributes seems comparable with supervised methods (i.e. GCN, GAT and GraphSAGE). 
The scale analysis in Fig. 7(b) reveals strong stability at $\mu = 0.1$, the model maintains highly consistent performance (0.80 to 0.64), indicating robust behavior under mild perturbations.
For edge manipulation scenarios in Figs. 7(c) and 7(d) DMoN exhibits robust behavior. Under random edge deletions, performance at $\mu = 0.1$ starts at a high 0.82 and gradually declines to 0.44, maintaining better overall performance compared to other methods. Similarly, targeted edge deletions begin at 0.80 and decline to 0.43, indicating that DMoN's spectral-based approach provides strong protection against both types of structural attacks.
A noteworthy observation is DMoN's consistent performance across both random and targeted edge deletions, particularly at lower $\mu$ values (0.1-0.3). The similar degradation patterns between random (0.82 to 0.44) and targeted (0.80 to 0.43) attacks at $\mu = 0.1$ suggest that DMoN's spectral and modularity based architecture provides balanced robustness against different types of structural perturbations. However, at $\mu = 0.5$, there's a slightly larger gap between random (0.28 final accuracy) and targeted (0.26 final accuracy) attack performance.
Overall, DMoN demonstrates better robustness compared to other pooling methods in specific perturbation scenarios, maintaining higher absolute performance values across all scenarios and showing strong robustness to both attribute perturbations and structural attacks. DMoN's modularity-based optimization that considers global network structure rather than local feature patterns, provides a grounded justification for its robustness since small changes in individual edges or attributes would have minimal impact on overall modularity scores. In addition, unsupervised graph learning combined with community level aggregation helps it learn global graph structure while simultaneously protecting against node level perturbations by averaging their effects.

\subsection{\label{sec:emp_summary}Summary of empirical study}

Here we present average stepwise differences in performance for GNN architectures, quantifying the gradual nature of ECS degradation under various perturbations to get a scope of overall robustness. These values were calculated by measuring the average absolute differences between consecutive steps as follows: 

1. For average drop across $\mu_i$ values for a specific model and perturbation type, we compute:

$$
\overline{\Delta\text{ECS}}_{\mu_i} = \frac{1}{|\mathcal{M}_i|} \sum_{\mathcal{M}_i}\frac{|\text{ECS}_{\text{initial}}(\mu_i) - \text{ECS}_{\text{final}}(\mu_i)|}{\text{ECS}_{\text{initial}}(\mu_i)}\times 100\%
,$$
where $\mathcal{M}_i$ is the set of all perturbations levels for all values $\mu_i$.

2. For average drop across perturbation types for a specific model, we compute.

$$
\overline{\Delta\text{ECS}}_{\text{pert}} = \frac{1}{|\mathcal{P}|} \sum_{p \in \mathcal{P}} \frac{|\text{ECS}_{\text{initial}}(p) - \text{ECS}_{\text{final}}(p)|}{\text{ECS}_{\text{initial}}(p)} \times 100\%
,$$
where $\mathcal{P}$ is the set of all perturbation types (mean, scale, random edge, targeted edge).
Table~\ref{tab:stepwise_diffs_by_mu} shows results averaged for different levels of community mixing parameters $\mu$ and  Table~\ref{tab:stepwise_diffs_by_perturb} reports the averages for perturbation types. Lower percentages indicate more gradual degradation and thus greater robustness. 

\begin{table*}[!ht]
\centering
\begin{tabular}{lcccccc}
\toprule
\textbf{Type} & \textbf{Model} & \textbf{$\mu = 0.1$} & \textbf{$\mu = 0.2$} & \textbf{$\mu =0.3$} & \textbf{$\mu = 0.4$} & \textbf{$\mu = 0.5$} \\
\midrule
\multirow{3}{*}{Supervised} & GCN & 8.28\% & 9.04\% & 8.62\% & 8.19\% & 7.51\% \\
 & GAT & 8.56\% & 8.47\% & 7.22\% & 5.91\% & 4.42\% \\
 & SAGE & 8.74\% & 8.15\% & 6.83\% & 5.26\% & 4.33\% \\
\midrule
\multirow{3}{*}{Unsupervised} 
& MinCut & 5.08\% & 5.43\% & 5.33\% & 3.35\% & 3.54\% \\
& DiffPool & 3.32\% & 3.60\% & 3.54\% & 1.81\% & 2.23\% \\
& DMoN & 5.86\% & 4.45\% & 4.49\% & 2.81\% & 2.67\% \\
\bottomrule
\end{tabular}
\caption{Average stepwise differences by model and community mixing parameter $\mu$. The lower the scores, more robust is the model. However, we evaluate in favor of a balance between performance and robustness.}
\label{tab:stepwise_diffs_by_mu}
\end{table*}

\begin{table*}[!ht]
\centering
\addtolength{\tabcolsep}{10pt}
\begin{tabular}{lccccc@{\hspace{0.5cm}}p{5cm}}
\toprule
\textbf{Type} & \textbf{Model} & \textbf{Location} & \textbf{Scale} & \textbf{Random Edge} & \textbf{Targeted Edge} \\
\midrule
\multirow{3}{*}{Supervised} & GCN & 13.80\% & 3.44\% & 8.18\% & 7.93\% \\
 & GAT & 10.49\% & 5.29\% & 6.02\% & 5.81\% \\
 & SAGE & 10.21\% & 4.72\% & 5.97\% & 5.73\% \\
\midrule
\multirow{3}{*}{Unsupervised} 
& MinCut & 7.33\% & 2.89\% & 4.08\% & 4.04\% \\
& DiffPool & 5.34\% & 2.65\% & 1.93\% & 1.92\% \\
& DMoN & 6.20\% & 2.88\% & 3.58\% & 3.54\% \\
\bottomrule
\end{tabular}
\caption{Average stepwise differences by model and perturbation type (averaged across all $\mu$ values).}
\label{tab:stepwise_diffs_by_perturb}
\end{table*}

From Table~\ref{tab:stepwise_diffs_by_mu} we notice that supervised GNN architectures (i.e., GCN, GAT, SAGE) exhibit larger stepwise differences (7.5-9.0\%, 4.4-8.6\%, and 4.3-8.7\% respectively), indicating more abrupt performance changes under perturbations. In contrast, approaches with unsupervised architectures that learn graph level representations (i.e., DiffPool, MinCut, DMoN) demonstrate notably smaller stepwise differences. DiffPool shows the most consistent and smallest stepwise differences (1.8-3.6\%), exhibiting more robust performance for community recovery. However these methods may have a lower overall performance in community recovery. DMoN pooling shows a balance between performance and robustness. Table~\ref{tab:stepwise_diffs_by_perturb} demonstrates how different perturbation types affect the robustness of community recovery for different GNNs. Mean perturbations consistently cause the largest stepwise differences across all architectures (ranging from 5.3\% for DiffPool to 13.8\% for GCN), indicating that causing a shift in location of distribution of node attributes can be adversarial towards community recovery. Scale perturbations, by contrast, produce more gradual degradation (2.7-5.3\%). Edge-based perturbations (both random and targeted) affect architectures similarly within each architecture, with DiffPool demonstrating remarkable stability against edge manipulations. However, taking into account overall community recovery, DMoN pooling seems to offer a good trade off. It shows modest ECS degradation at even strong community mixing ($\mu = 0.5$) with 2.67\% average stepwise drop, but at same time is competitive in performance to supervised methods.  This makes DMoN a robust option as it achieves a balance between robustness and community recovery. These results suggest that DMoN's optimization of the modularity measure creates inherent robustness to structural attacks while maintaining competitive performance, making it potentially more suitable than DiffPool for applications where both robustness and high accuracy are required.

\subsection{\label{sec:adversarial}Advanced graph perturbations}

\subsubsection{\label{sec:level3}Nettack}

\begin{table*}
\centering
\begin{tabular}{l|l|cc|cc}
\hline
\multirow{2}{*}{\bf Type} & \multirow{2}{*}{\bf Method} & \multicolumn{2}{c}{\bf Nettack} & \multicolumn{2}{c}{\bf Metattack} \\
\cline{3-6}
& & $\mu = 0.1$ & $\mu = 0.5$ & $\mu = 0.1$ & $\mu = 0.5$ \\
\hline
\multirow{3}{*}{Supervised} & GCN  & 0.53 ($\downarrow 0.27) $ & 0.28 ($\downarrow 0.21)$ & 0.58 ($\downarrow 0.22) $ &  0.28 ($\downarrow 0.11)$ \\
& GAT & 0.57 ($\downarrow 0.20) $ & 0.32 ($\downarrow 0.12)$ & 0.63 ($\downarrow 0.15) $ &  0.34 ($\downarrow 0.08)$  \\
& SAGE & 0.49 ($\downarrow 0.29) $ & 0.34 ($\downarrow 0.14)$ & 0.54 ($\downarrow 0.24) $ &  0.29 ($\downarrow 0.14)$ \\
\hline
\multirow{3}{*}{Unsupervised} & MinCut & 0.44 ($\downarrow 0.10) $ & 0.35 ($\downarrow 0.08)$ & 0.51 ($\downarrow 0.06) $ &  0.27 ($\downarrow 0.12)$ \\
& DiffPool & 0.43 ($\downarrow 0.12) $ & 0.19 ($\downarrow 0.24)$ & 0.43 ($\downarrow 0.09) $ &  0.21 ($\downarrow 0.11)$ \\
& DMoN & 0.64 ($\downarrow 0.18) $ & 0.46 ($\downarrow 0.09)$ & 0.69 ($\downarrow 0.14) $ &  0.33 ($\downarrow 0.11)$ \\
\hline
\end{tabular}
\caption{ECS scores for comparison of Nettack and Metattack for community structures with $\mu = 0.1,0.5$.}
\label{tab:adversarial results}
\end{table*}

Network perturbation attacks, specifically Nettack and Metattack, represent sophisticated approaches to compromising graph-based learning systems through structural perturbations. Nettack operates as a targeted attack mechanism that modifies the graph structure around specific nodes by strategically adding or removing edges and manipulating node features, while maintaining key graph properties to ensure perturbations remain undetectable. Metattack, in contrast, functions as a meta-learning based attack that aims to deteriorate the overall model performance by learning generalizable adversarial perturbations that can transfer across different graph neural network architectures.
The experimental results demonstrate varying levels of robustness across different graph-based community detection methods when controlled for Nettack and Metattack under different community strengths ($\mu$). Table~\ref{tab:adversarial results} shows that DMoN architecture exhibits the strongest baseline performance, achieving ECS scores of 0.64 and 0.69 for LFR network with $\mu$ = 0.1 for Nettack and Metattack. However, its performance significantly degrades under weak community strength ($\mu$ = 0.5), falling to 0.46 and 0.33, representing relative decreases of 0.09 and 0.11 respectively. GAT architecture demonstrates notable robustness compared to other approaches, with a smaller deterioration of ECS scores of 0.57 and 0.63. More importantly, GAT shows the smallest performance degradation under Metattack ($\downarrow$0.08 at $\mu$ = 0.5), suggesting that the attention mechanism helps maintain structural integrity even under aggressive perturbation attempts.

GCN  and SAGE show similar robustness patterns, with significant performance drops under both attack types. GCN's accuracy decreases by $\downarrow$0.27 and $\downarrow$0.22  respectively, while SAGE experiences drops of $\downarrow$0.29 and $\downarrow$0.24 for Nettack and Metattack respectively. This suggests that supervised architectures may be more susceptible to structural perturbation attacks.

The hierarchical DiffPool approach, while showing moderate stability under Metattack ($\downarrow$0.09 at $\mu$ = 0.1), exhibits the poorest overall performance with accuracies dropping to 0.19 and 0.21 under high perturbation rates. This indicates that hierarchical pooling operations may amplify the effects of structural perturbations. Conversely, the MinCut method demonstrates the most consistent performance across perturbation rates, with the smallest average degradation under Nettack ($\downarrow$0.10 and $\downarrow$0.08), though its baseline performance is relatively lower than other methods.

Overall, the results suggest that attention-based mechanisms and spectral approaches such as DMoN offer better protection against perturbation attacks, while methods incorporating edge-level attention such as GAT might exhibit more  robustness under more aggressive attack scenarios. The significant performance degradation across all methods for weaker community structures ($\mu$ = 0.5) highlights important challenge of robust community detection algorithms using deep graph encoders.

\subsection{\label{sec:citation_net}Real-world citation networks}

\begin{figure}[!htb]
    \centering
    \includegraphics[width=0.48\textwidth]{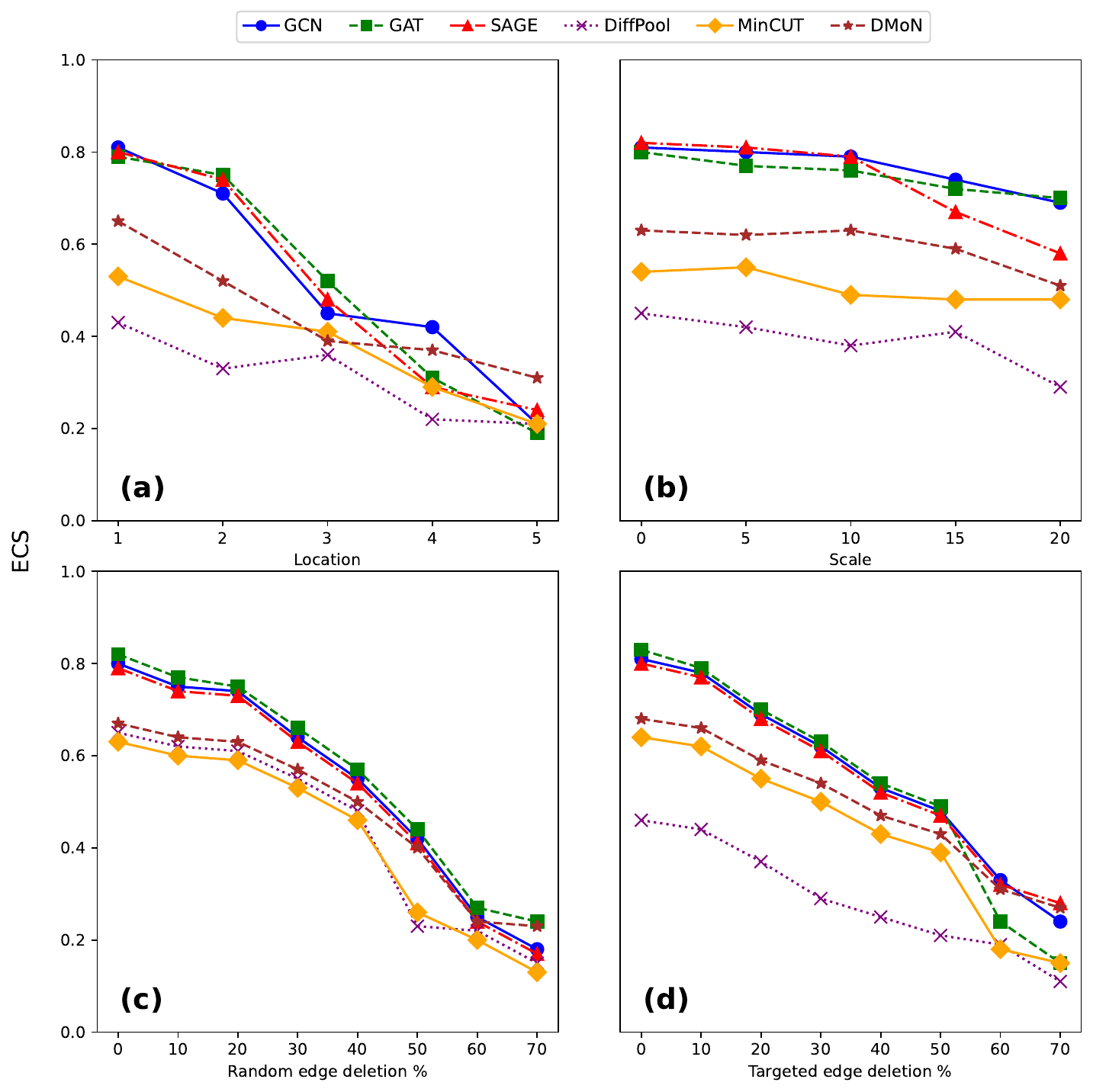}
    \caption{ECS scores for Cora dataset. Under node attribute perturbations, (a) shows ECS score across different means and (b) represents scale perturbations. Under structural perturbations, (c) shows results for random edge perturbations and (d) for targeted attacks based on betweenness centrality.}
    \label{Cora}
\end{figure}

Fig.~\ref{Cora} shows the result on robustness for the Cora dataset. We observe significant differences in community recovery capabilities between supervised (GCN, GAT, GraphSAGE) and unsupervised (DiffPool, MinCut, DMoN) GNN architectures as measured by ECS. Supervised approaches consistently demonstrate high community detection performance, with GCN achieving the highest baseline ECS of 0.81. Under mean perturbation, supervised architectures experience substantial ECS drops, with GCN declining by 74.1\% (from 0.81 to 0.21), GAT by 76.0\% (from 0.79 to 0.19), and GraphSAGE by 70.0\% (from 0.80 to 0.24). Unsupervised methods generally demonstrate lower baseline ECS but similar relative declines specifically, DMoN drops by 52.3\% (from 0.65 to 0.31), while MinCut decreases by 60.4\% (from 0.53 to 0.21). Under targeted edge attacks at the most severe level, supervised architectures show significant degradation, with GCN dropping to 0.24 (70.4\% decrease), GAT to 0.15 (81.9\% decrease), and GraphSAGE to 0.28 (65.0\% decrease). Comparatively, DMoN, the best performing unsupervised model, maintains an ECS of 0.27 (60.3\% decrease).

\begin{figure}[!htb]
    \centering
    \includegraphics[width=0.48\textwidth]{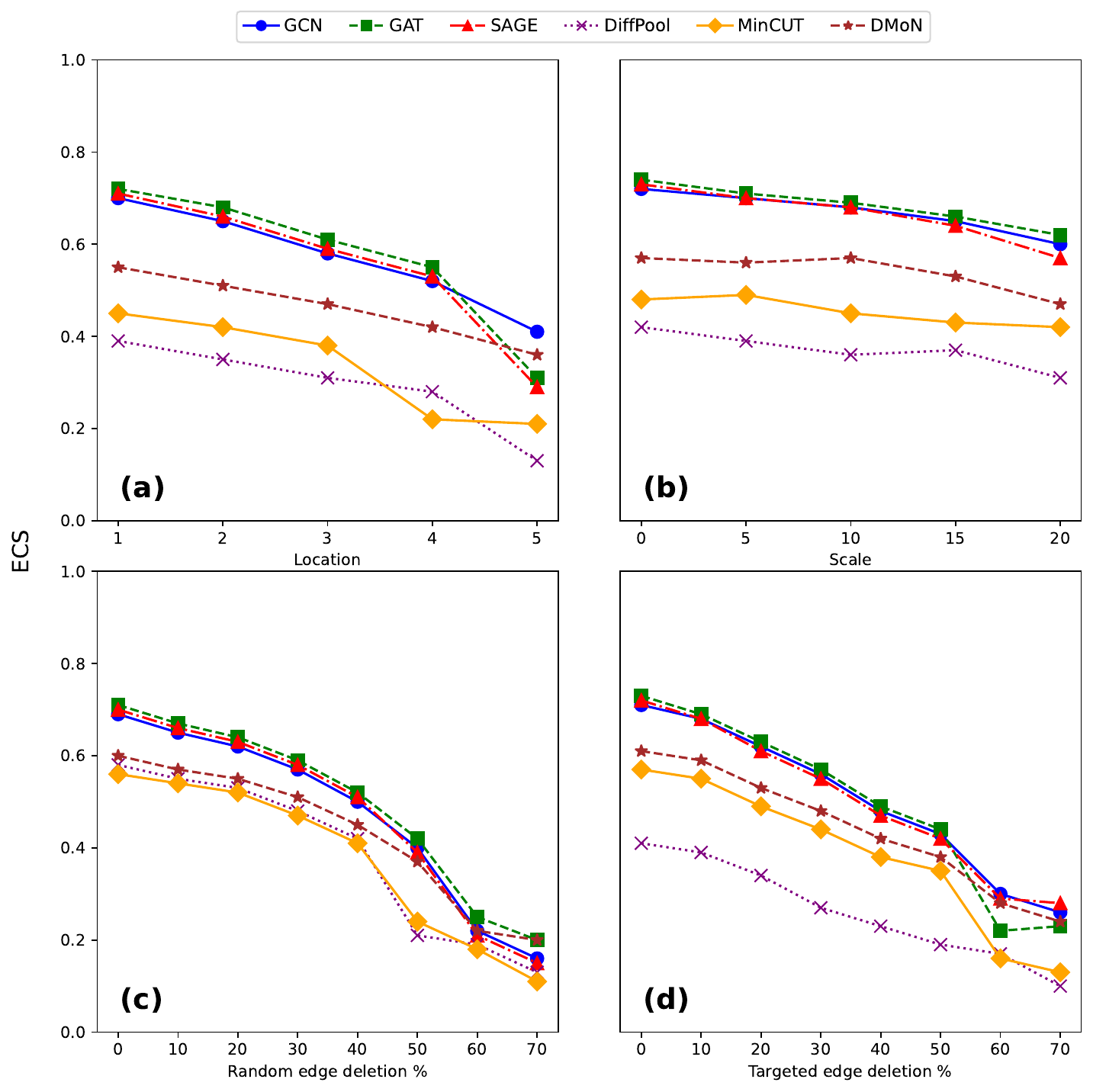}
    \caption{ECS scores for Citeseer dataset. Under node attribute perturbations, (a) shows ECS score across different means and (b) represents scale perturbations. Under structural perturbations, (c) shows results for random edge perturbations and (d) for targeted attacks based on betweenness centrality.}
    \label{cite}
\end{figure}

Fig. ~\ref{cite} shows results for robustness analysis on Citeseer dataset. For the Citeseer dataset, GAT achieves the strongest baseline performance among all architectures, with supervised methods generally outperforming unsupervised approaches where strong community structure is present. However, GAT demonstrates a higher vulnerability across all types of perturbations. Under location perturbations, GAT experiences severe degradation with a 65.33\% average performance drop the, highest among all architectures tested. For location perturbations, GAT shows relatively better stability with a 13.49\% drop, but under structural attacks, its lack of robustness becomes pronounced with 70.01\% degradation under random edge perturbations and 73.95\% under targeted attacks. In contrast, DMoN exhibits strong robustness among unsupervised approaches with only a 34.28\% drop under location perturbations and demonstrates the lowest vulnerability to targeted attacks at 50.27\%. Under extreme perturbations, DMoN shows better robustness surpassing supervised methods. 

\begin{figure}[!htb]
    \centering
    \includegraphics[width=0.48\textwidth]{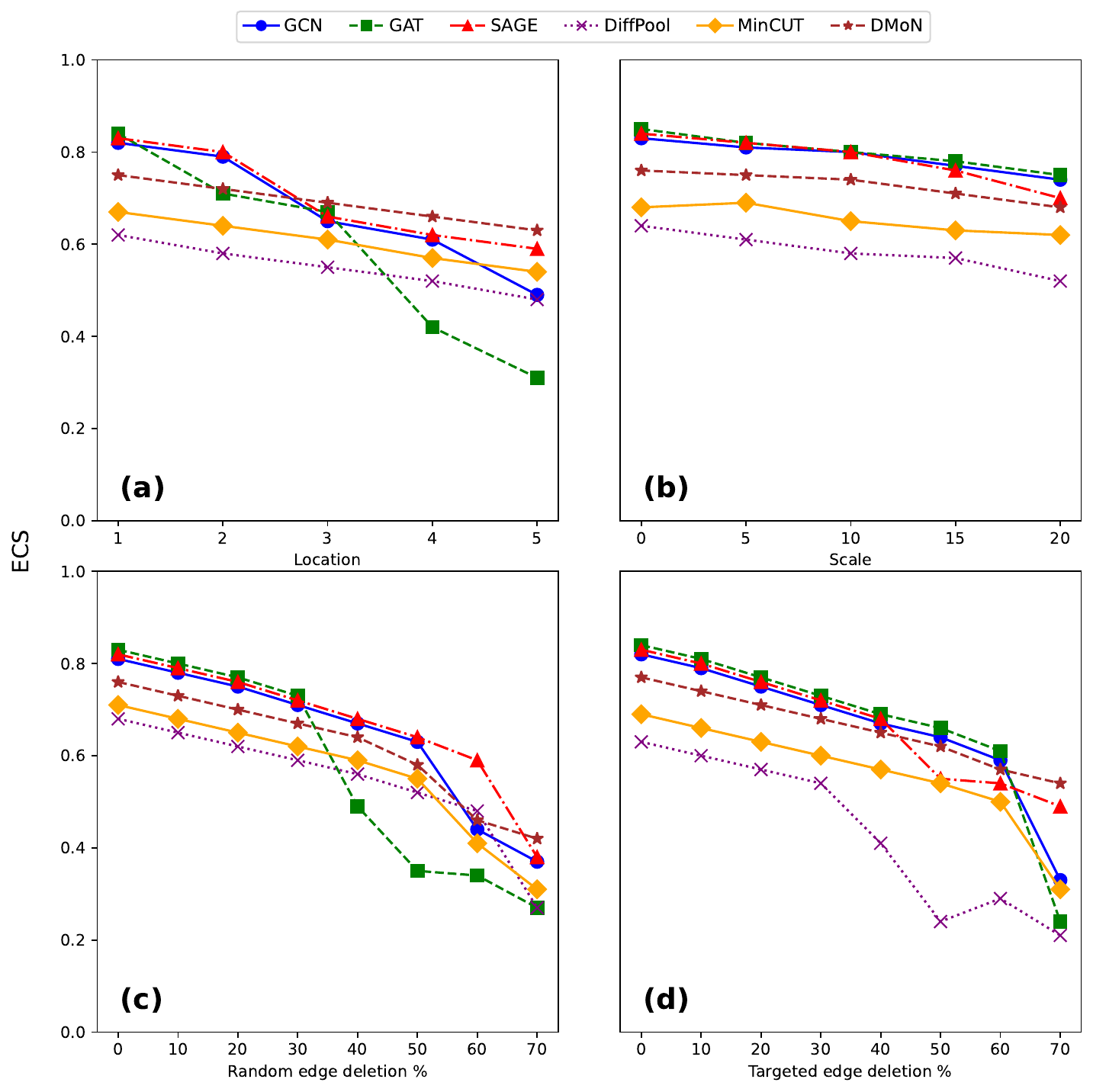}
    \caption{ECS scores for PubMed biomedical dataset. Under node attribute perturbations, (a) shows ECS score across different means and (b) represents scale perturbations. Under structural perturbations, (c) shows results for random edge perturbations and (d) for targeted attacks based on betweenness centrality.}
    \label{fig:pubmed}
\end{figure}

Fig. ~\ref{fig:pubmed} shows robustness analysis on Pubmed dataset. We observe that while GAT initially achieves the highest community detection performance with an ECS of 0.84, it demonstrates vulnerability under perturbation, dropping to 0.31 (a 63.1\% decline) under mean perturbation and to just 0.24 under severe targeted attacks. In stark contrast, DMoN emerges robust across all test conditions, with only a 16\% drop for mean perturbations. DMoN retains also an ECS of 0.54 under targeted. GraphSAGE also demonstrates surprising robustness for a supervised model, maintaining 0.59 ECS under severe mean perturbation. The PubMed results suggest that the graph structure and domain characteristics may significantly influence architecture behavior. DMoN shows substantial advantages in maintaining the community detection capability under adversarial conditions. 

We summarize our findings on real world networks in Table~\ref{tab:dataset_drops_by_perturbation} across Cora, CitSeer, and PubMed datasets. For supervised architectures, GCN provides a good balance of performance and robustness, while GAT achieves the highest baseline scores (peaking at 0.84 on PubMed) but suffers dramatic degradation under perturbation. Among unsupervised approaches, DMoN consistently demonstrates robustness despite the fact that overall performance is lower compared to supervised methods. However, under stronger perturbations, DMoN and unsupervised community detection seem to be more robust suggesting metrics such as modularity maybe key to community recovery when graph integrity is challenged. 
\begin{table*}[!ht]
\centering
\begin{tabular}{llcccc}
\toprule
\textbf{Type} & \textbf{Model} & \textbf{Location} & \textbf{Scale} & \textbf{Random Edge} & \textbf{Targeted Edge} \\
\midrule
\multirow{3}{*}{Supervised} & GCN & 51.92\% & 14.11\% & 69.54\% & 64.50\% \\
& GAT & 65.33\% & 13.49\% & 70.01\% & 73.95\% \\
& SAGE & 52.69\% & 22.62\% & 70.24\% & 55.69\% \\
\midrule
\multirow{3}{*}{Unsupervised} & MinCut & 44.37\% & 10.81\% & 72.02\% & 69.61\% \\
& DiffPool & 46.80\% & 26.83\% & 71.60\% & 72.79\% \\
& DMoN & 34.28\% & 15.71\% & 59.03\% & 50.27\% \\
\bottomrule
\end{tabular}
\caption{Average percentage drops by model and perturbation type (averaged across datasets).}
\label{tab:dataset_drops_by_perturbation}
\end{table*}


\section{\label{sec:level2}Conclusion}

Our comprehensive analysis of GNN-based community detection robustness has revealed a number of important phenomena. First, we have found that that while supervised approaches (GCN, GAT, GraphSAGE) achieve higher baseline performance on well-structured data, unsupervised methods, particularly DMoN with its modularity based objective, tend to demonstrate superior robustness under severe perturbations. This robustness gap is especially pronounced in adversarial settings. Community strength (mixing parameter $\mu$) appears to be a critical factor in determining robustness, with stronger communities providing substantial protection against all perturbation types. In light of these findings, we envision the following promising research directions.
First,  investigating neural scaling laws for GNNs in community detection tasks may be essential for designing new robust models. While scaling laws have been extensively studied for language and vision models, their application to graph learning remains largely unexplored. Understanding how community detection performance scales with model size, training data volume, and computational resources could provide critical insights for architecture design. Second, it appears that
looking at GNNs through the lens of hyperbolic embeddings~\citep{budel2024random, desy2023dimension, yang2022hyperbolic} may be an attractive path to shed new light on their ability to sustain various uncertainties and perturbations not only for community detection but more general downstream tasks.

Third, it will be  important to examine whether GNNs follow power-law relationships between these factors and model performance, particularly, in the context of robustness to perturbations. This could help establish more principled approaches to model selection and training that balance computational efficiency with perturbation resistance in community detection applications.
\section*{Acknowledgments}

This paper has been coauthored by UT-Battelle, LLC under Contract No.\ DE-AC05-00OR22725 with the U.S.\ Department of Energy. The publisher, by accepting the article for publication, acknowledges that the U.S.\ government retains a nonexclusive, paid up, irrevocable, world-wide license to publish or reproduce the published form of the manuscript, or allow others to do so, for U.S.\ government purposes. The DOE will provide public access to these results in accordance with the DOE Public Access Plan (http://energy.gov/downloads/doe-public-access-plan). This material is based upon work supported by the U.S. Department of Energy, Office of Science, Office of Advanced Scientific Computing Research under Contract No. DE-AC05-00OR22725.

\clearpage
\bibliography{ref.bib}

\clearpage

\section*{Appendix}

\subsection{\label{app-A} Theoretical details}

\subsubsection*{Cut-based metrics} Let $\{\mathcal{C}_i\}_{i=1}^K$ be disjoint partitions of $\mathcal{V}$. We can define a cut as
$$
cut(\mathcal{C}_1, \ldots, \mathcal{C}_K) = \frac{1}{2}\sum_{k=1}^K |(u, v) \in E : u \in \mathcal{C}_k, v \in \hat{\mathcal{C}}_k|,
$$
here $\hat{\mathcal{C}}_k$ is the complement of partition $\mathcal{C}$. A cut defines how many edges are present within community boundaries and cut based metrics aim to minimize the number of edges between clusters while balancing cluster sizes. To ensure clustering algorithms don't degenerate into trivial clusters of singular nodes, there are several methods such as normalized cuts and ratio cuts that have been introduced that take into account partition sizes. However, real networks often lack good cuts, as nodes may participate in multiple clusters simultaneously. MinCutPool adapts the notion of normalized cut for use as a regularizer in GNN in order to treat spectral clustering as differentiable pooling strategy. 

\subsubsection*{Modularity} This approaches clustering from a statistical perspective, measuring the deviation of suggested clustering from what clustering would like for random graph. Modularity $Q$ is formally defined as:
\[
Q = \frac{1}{2m}\sum_{ij}\left[\mathbf{A}_{ij} - \frac{d_id_j}{2m}\right]\delta(c_i, c_j)
\],
where $\delta(c_i, c_j) = 1$ if $i$ and $j$ are in the same cluster and 0 otherwise. Modularity can be reformulated as a matrix trace:
\[
Q = \frac{1}{2m} \text{Tr}(\mathbf{C}^T\mathbf{B}\mathbf{C})
\],
where $\mathbf{C} \in {0, 1}^{n\times k}$ is the cluster assignment matrix and $\mathbf{B} = \mathbf{A} - \frac{dd^T}{2m}$ is the modularity matrix.

\subsection{ECS technical summary}
\label{app-B}
The foundation lies in constructing a bipartite graph $\mathcal{B}(V \cup C, E)$ with vertex set $V$, $C$ represents clusters, and $ E $ defines edges between elements and their clusters. When dealing with hierarchical structures, each cluster $c_\beta \in C$ is assigned a level $l_\beta \in [0,1]$, with edge weights determined by a hierarchical weighting function:
$$h(l_\beta) = e^{rl_\beta}$$
where $r$ acts as a scaling parameter controlling hierarchical influence. The method then projects this structure onto an element graph through weighted edges, capturing the relationships induced by common cluster memberships. Edge weights $w_{ij}$ between elements are calculated by normalizing shared cluster affiliations:
$$w_{ij} = \frac{\sum_\gamma a_{i\gamma}a_{j\gamma}}{\sum_\gamma \sum_m a_{i\gamma}a_{m\gamma}}$$
where $a_{i\gamma}$ represents entries in the bipartite adjacency matrix. The element-wise similarity between clusterings is then computed using personalized PageRank with a stationary distribution $p_i = (1.0-\alpha)v_i + \alpha v_i\mathbf{W}$, where $\alpha$ controls the influence of overlapping and hierarchical structures. The final similarity score between two clusterings $\mathbf{C}$ and $\mathbf{C}'$ is calculated as the average of element-wise similarities:
$$S(\mathcal{C},\mathcal{C}') = 1.0 - \frac{1}{2\alpha N}\sum_{i=1}^N\sum_{j=1}^N |p_{ij}^{\mathcal{C}} - p_{ij}^{\mathcal{C}'}|$$
This formulation effectively captures community structure by incorporating multiple aspects including overlapping communities represented through multiple edges in the affiliation graph, hierarchical structure preserved through weighted edges, and both local and global network topology maintained through the personalized PageRank diffusion process. This approach overcomes traditional biases present in other similarity measures while providing a unified framework for comparing disjoint, overlapping, and hierarchically structured communities.

\subsection{\label{app-C} Additional Experiments}

In this section we analyze GNN architecture performance on ADC-SBM.
Based on the experimental data across six GNN models tested on ADC-SBM~\citep{tsitsulin2022synthetic} benchmarks for community detection performance, there is a clear robustness hierarchy in terms of community recovery capabilities. ADC-SBM extends the classical stochastic block model to incorporate both node attributes and degree heterogeneity, with parameters ranging from $\rho = (0.1,0.5)$ representing increasing levels of community strength. From Figs 11-16, our additional experiments  on ADC-SBM are consistent with our main finding that DMoN maintains the best community detection performance across varying community strengths. From Tables 8 and 9 we observe that supervised methods show more abrupt performance degradation with larger stepwise differences (6-7.8\% for GAT vs. 2.7-6.4\% for DMoN). Congruent with observations in LFR experiments, DMoN's robustness relies on its modularity-based optimization which provides stability against perturbations, as modularity creates inherent robustness to structural attacks. In Table 10, we observe that under adversarial attacks that is, Nettack and Metattack, DMoN demonstrates strong robustness as well with minimal degradation for both levels of community complexity ($\rho=0.1$ and $\rho=0.5$).
\subsection{\label{app-D} Hyper-parameter details}
Following established practices in graph neural network research for community detection~\citep{kipf2016semi, tsitsulin2023graph, bianchi2020spectral_mincut, ying2018hierarchical_diffpool}, we adopt standardized hyperparameter settings to ensure fair comparison across all methods. Our hyperparameter selection is based on commonly reported values in the literature and empirical validation on benchmark datasets.

For supervised methods (GCN, GAT, GraphSAGE), we follow the configuration established by \cite{kipf2016semi,velivckovic2017gat,hamilton2017sage} with learning rate of 0.01. For GCN, we use 2 layers with 32 hidden units per layer. GAT employs 2 layers with 8 attention heads in the first layer and 1 attention head in the second layer, each with 8 features per head. GraphSAGE uses 2 layers with 128 hidden units each and mean aggregator function, following the original implementation guidelines~\citep{hamilton2017sage}.

For unsupervised methods (MinCUT, DiffPool, DMoN),
we employ single-layer GCN layer with 64 hidden units to learn representations followed by the respective pooling mechanisms. MinCUT is configured with orthogonality loss weight of 1.0 to encourage balanced cluster assignments. DMoN is trained with collapse regularization weight of 1.0, learning rate of 0.001, and 1000 training epochs. Following the configuration established by Tsitsulin et al.~\citep{tsitsulin2023graph} all methods use dropout rate of 0.1 for regularization during training and Adam optimizer.

\begin{figure}[H]
    \centering
\includegraphics[width=0.48\textwidth]{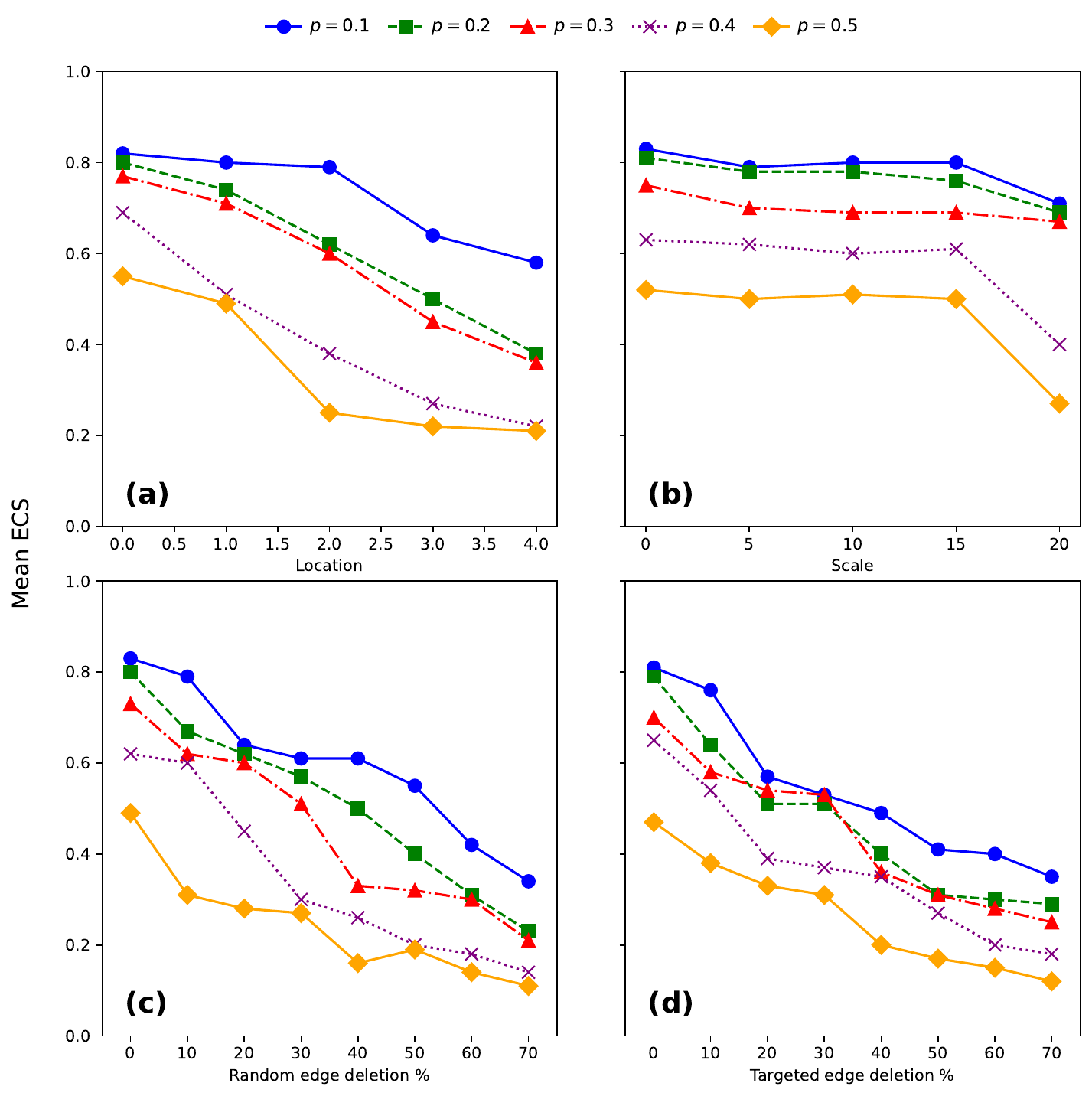}
    \caption{ECS scores for GCN-based community detection under four perturbation scenarios. Under node attribute perturbations, (a) shows ECS score across different means and (b) represents scale perturbations. Under structural perturbations, (c) shows results for random edge perturbations and (d) for targeted attacks based on betweenness centrality.}
    \label{fig:sbm_gcn}
\end{figure}

\begin{figure}
    \centering
    \includegraphics[width=0.48\textwidth]{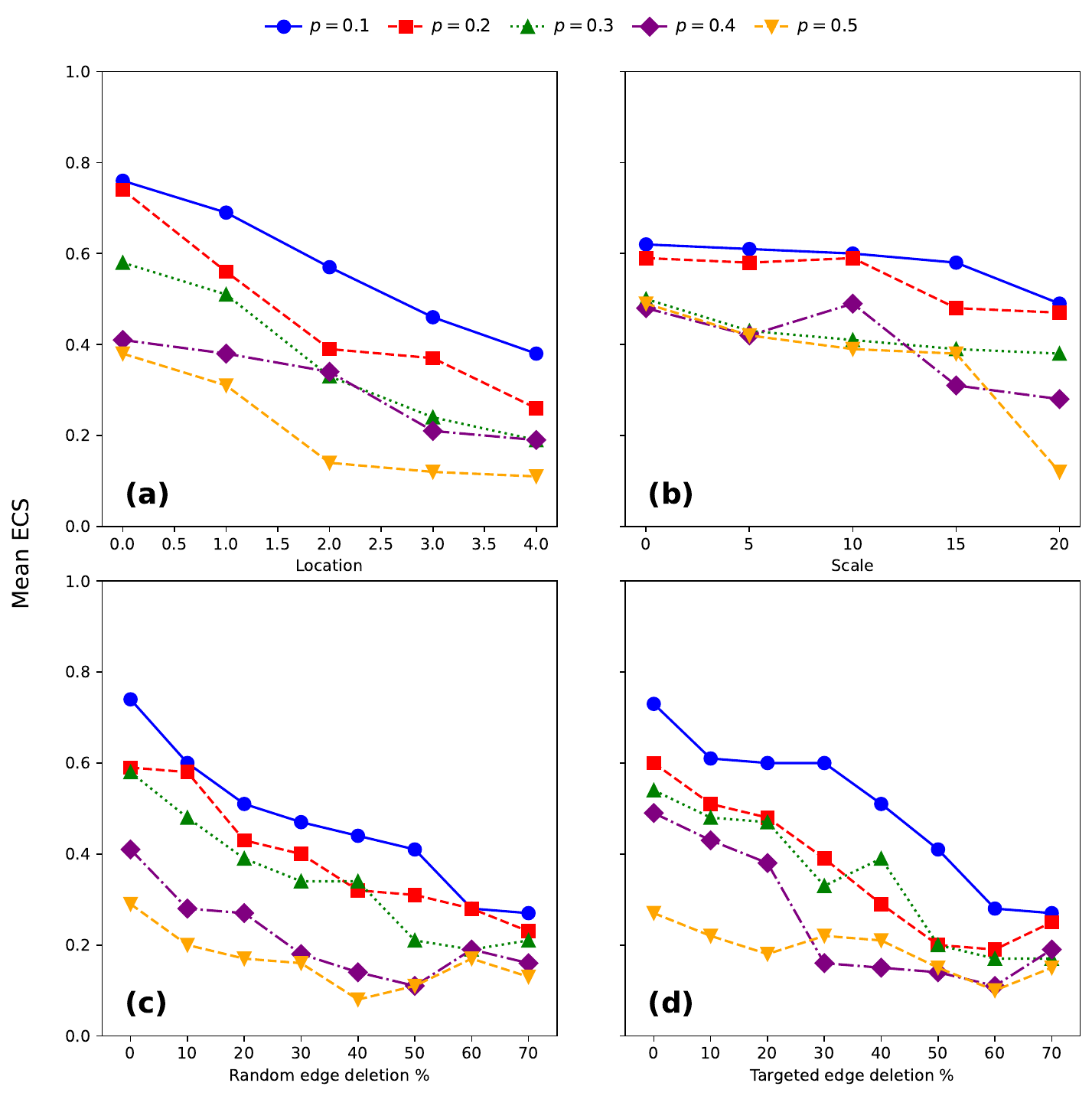}
    \caption{ECS scores for GAT-based community detection under four perturbation scenarios. Under node attribute perturbations, (a) shows ECS score across different means and (b) represents scale perturbations. Under structural perturbations, (c) shows results for random edge perturbations and (d) for targeted attacks based on betweenness centrality.}
    \label{fig:sbm_gat}
\end{figure}

\begin{figure}
    \centering
    \includegraphics[width=0.48\textwidth]{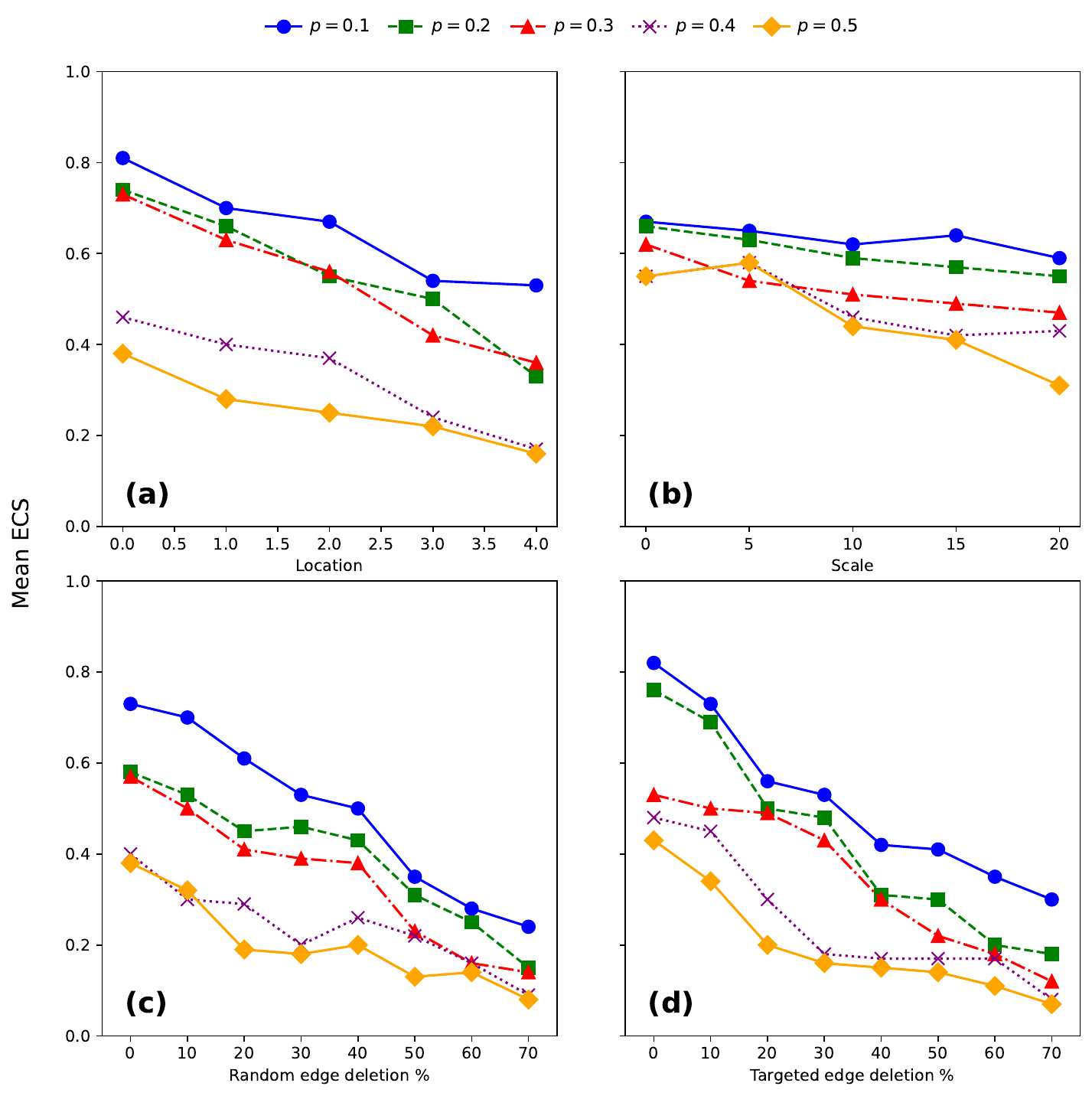}
    \caption{ECS scores for GraphSAGE-based community detection under four perturbation scenarios. Under node attribute perturbations, (a) shows ECS score across different means and (b) represents scale perturbations. Under structural perturbations, (c) shows results for random edge perturbations and (d) for targeted attacks based on betweenness centrality.}
    \label{fig:sbm_sage}
\end{figure}

\begin{figure}
    \centering
    \includegraphics[width=0.48\textwidth]{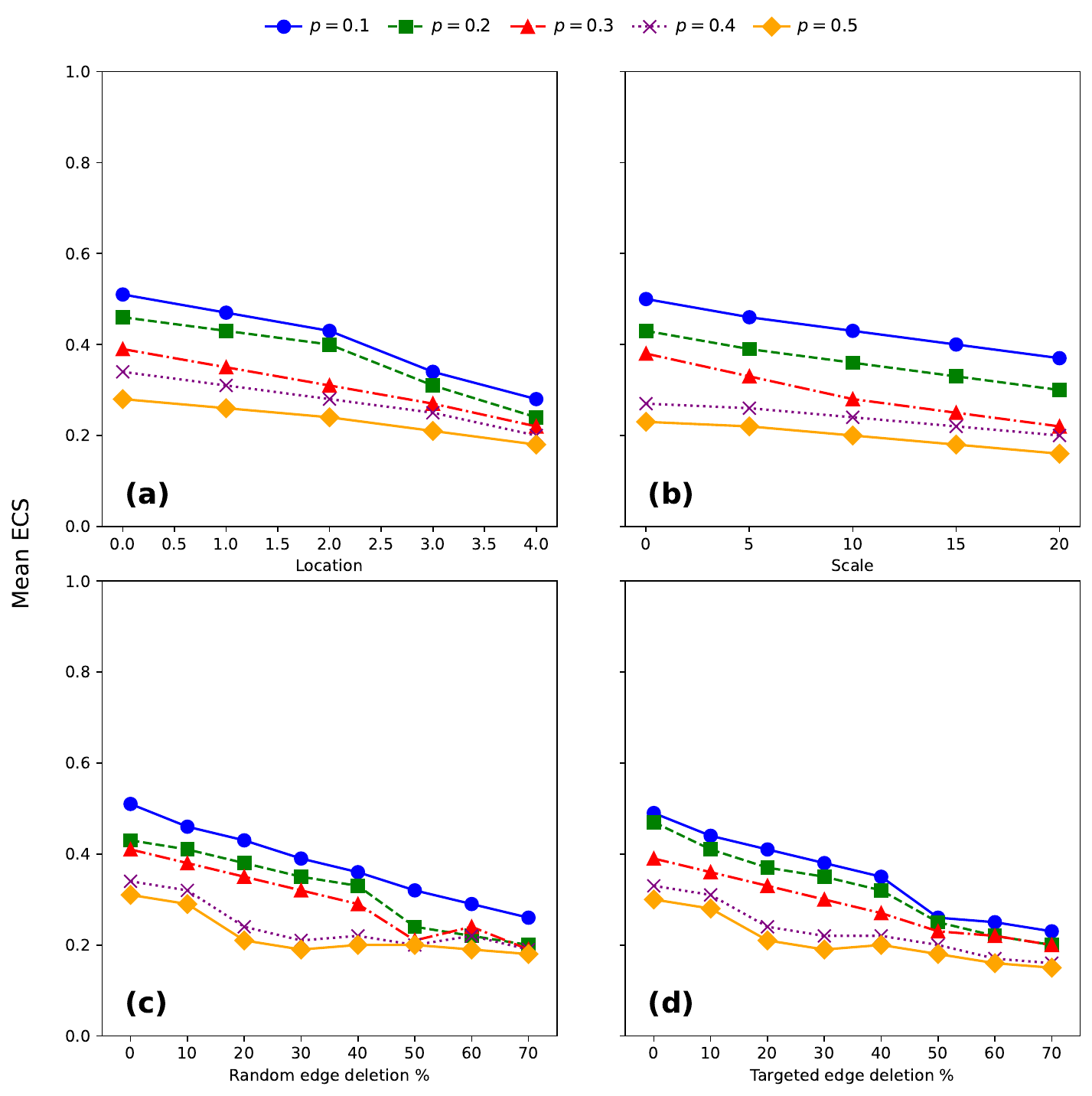}
    \caption{ECS scores for Diffpool-based community detection under four perturbation scenarios. Under node attribute perturbations, (a) shows ECS score across different means and (b) represents scale perturbations. Under structural perturbations, (c) shows results for random edge perturbations and (d) for targeted attacks based on betweenness centrality.}
    \label{fig:sbm_diffpool}
\end{figure}

\begin{figure}
    \centering
    \includegraphics[width=0.48\textwidth]{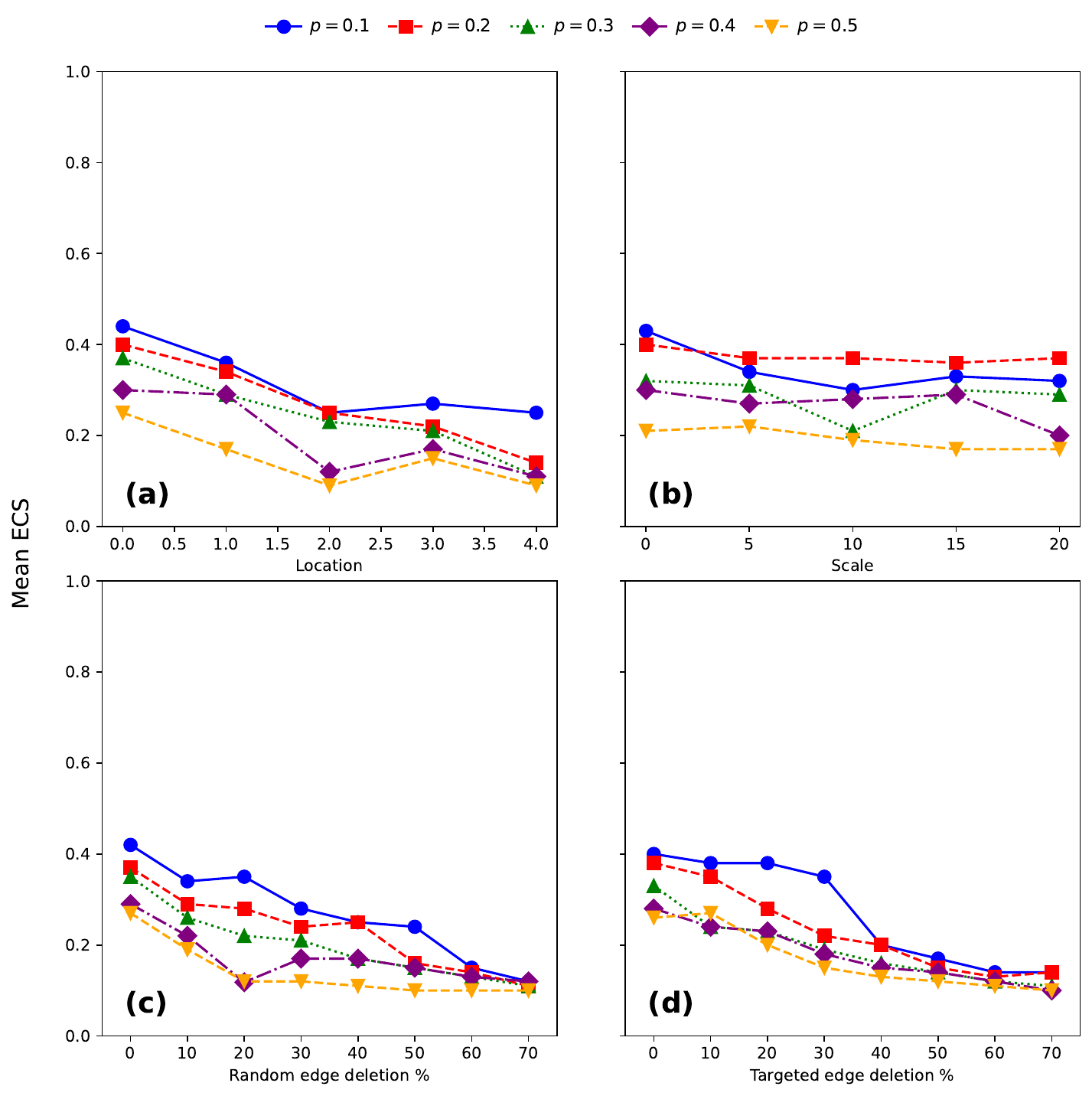}
    \caption{ECS scores for Mincut-based community detection under four perturbation scenarios. Under node attribute perturbations, (a) shows ECS score across different means and (b) represents scale perturbations. Under structural perturbations, (c) shows results for random edge perturbations and (d) for targeted attacks based on betweenness centrality.}
    \label{fig:sbm_mincut}
\end{figure}

\begin{figure}
    \centering
    \includegraphics[width=0.48\textwidth]{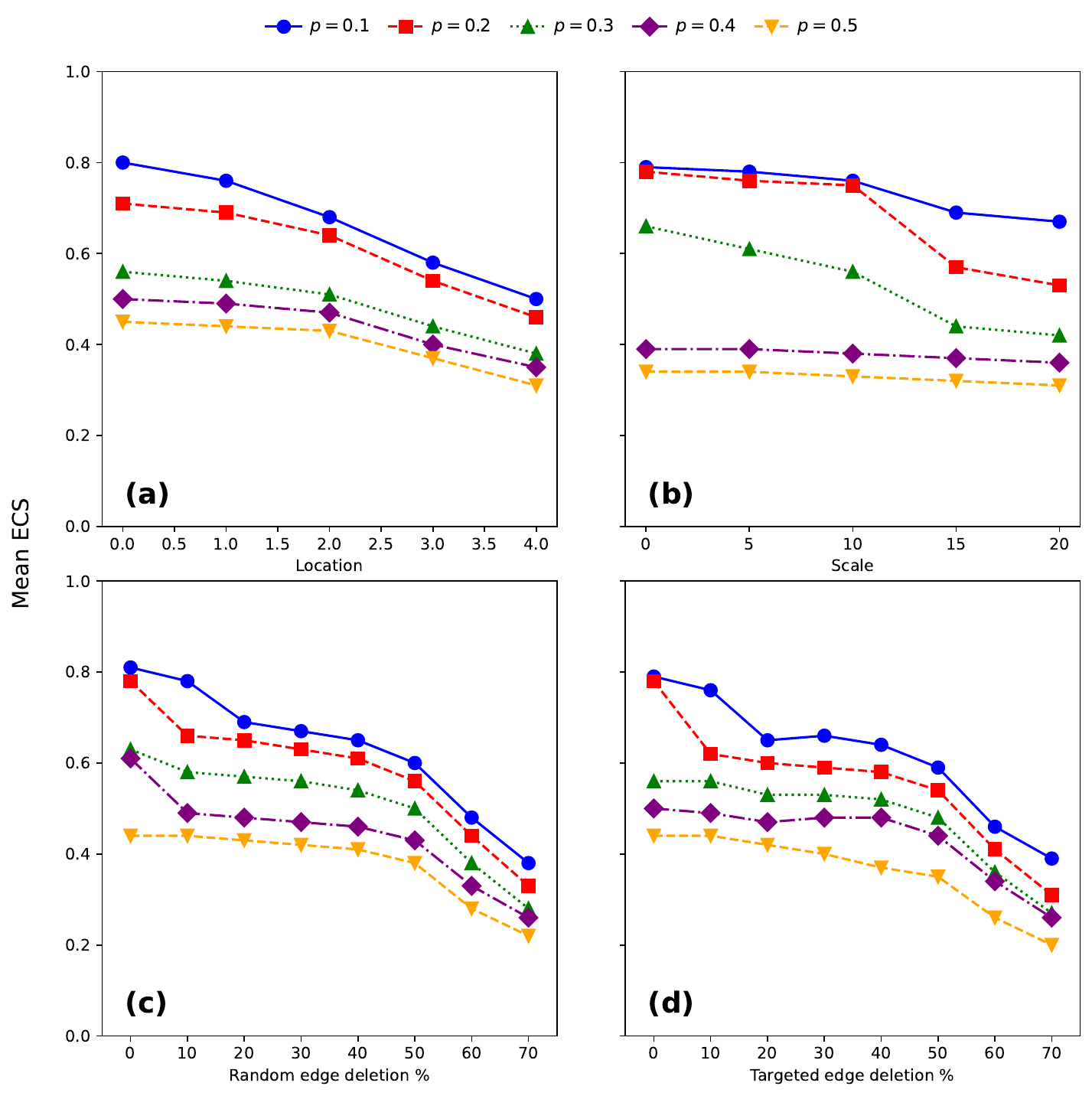}
    \caption{ECS scores for DMoN-based community detection under four perturbation scenarios. Under node attribute perturbations, (a) shows ECS score across different means and (b) represents scale perturbations. Under structural perturbations, (c) shows results for random edge perturbations and (d) for targeted attacks based on betweenness centrality.}
    \label{fig:sbm_dmon}
\end{figure}

\begin{table*}
\centering
\begin{tabular}{lcccccc}
\toprule
\textbf{Type} & \textbf{Model} & \textbf{$\rho = 0.1$} & \textbf{$\rho = 0.2$} & \textbf{$\rho =0.3$} & \textbf{$\rho = 0.4$} & \textbf{$\rho = 0.5$} \\
\midrule
\multirow{3}{*}{Supervised} & GCN & 5.66\% & 7.24\% & 6.52\% & 7.81\% & 6.25\% \\
 & GAT & 6.14\% & 7.84\% & 6.63\% & 7.27\% & 6.18\% \\
 & SAGE & 5.77\% & 6.91\% & 6.29\% & 5.10\% & 5.21\% \\
\midrule
\multirow{3}{*}{Unsupervised} 
& MinCut & 3.90\% & 3.64\% & 3.37\% & 3.08\% & 2.44\% \\
& DiffPool & 5.83\% & 6.91\% & 6.29\% & 5.06\% & 5.15\% \\
& DMoN & 5.64\% & 6.36\% & 4.90\% & 3.31\% & 2.73\% \\
\bottomrule
\end{tabular}
\caption{Average stepwise differences by model and community strength parameter $\rho$ - Averaged across all perturbations.}
\label{tab:8}
\end{table*}

\begin{table*}
\centering
\addtolength{\tabcolsep}{10pt}
\begin{tabular}{lccccc@{\hspace{0.5cm}}p{5cm}}
\toprule
\textbf{Type} & \textbf{Model} & \textbf{Location} & \textbf{Scale} & \textbf{Random Edge} & \textbf{Targeted Edge} \\
\midrule
\multirow{3}{*}{Supervised} & GCN & 9.37\% & 4.03\% & 6.98\% & 6.41\% \\
 & GAT & 10.21\% & 5.61\% & 7.92\% & 7.67\% \\
 & SAGE & 7.77\% & 3.53\% & 5.62\% & 6.48\% \\
\midrule
\multirow{3}{*}{Unsupervised} 
& MinCut & 5.27\% & 1.61\% & 3.30\% & 3.01\% \\
& DiffPool & 7.82\% & 3.48\% & 5.58\% & 6.52\% \\
& DMoN & 5.11\% & 3.40\% & 5.12\% & 4.67\% \\
\bottomrule
\end{tabular}
\caption{Average stepwise differences by model and perturbation type (averaged across all $\rho$ values).}
\label{tab:9}
\end{table*}

\begin{table*}
\centering
\begin{tabular}{l|l|cc|cc}
\hline
\multirow{2}{*}{\bf Type} & \multirow{2}{*}{\bf Method} & \multicolumn{2}{c}{\bf Nettack} & \multicolumn{2}{c}{\bf Metattack} \\
\cline{3-6}
& & $\rho = 0.1$ & $\rho = 0.5$ & $\rho= 0.1$ & $\rho = 0.5$ \\
\hline
\multirow{3}{*}{Supervised} & GCN  & 0.57 ($\downarrow 0.23$) & 0.36 ($\downarrow 0.20$) & 0.63 ($\downarrow 0.24$) &  0.30 ($\downarrow 0.13$) \\
& GAT & 0.67 ($\downarrow 0.14$) & 0.38 ($\downarrow 0.11$) & 0.69 ($\downarrow 0.14$) &  0.39 ($\downarrow 0.07$)  \\
& SAGE & 0.55 ($\downarrow 0.27$) & 0.40 ($\downarrow 0.13$) & 0.61 ($\downarrow 0.22$) &  0.35 ($\downarrow 0.12$) \\
\hline
\multirow{3}{*}{Unsupervised} & MinCut & 0.50 ($\downarrow 0.10$) & 0.41 ($\downarrow 0.08$) & 0.58 ($\downarrow 0.07$) &  0.33 ($\downarrow 0.11$) \\
& DiffPool & 0.49 ($\downarrow 0.12$) & 0.25 ($\downarrow 0.23$) & 0.50 ($\downarrow 0.09$) &  0.27 ($\downarrow 0.10$) \\
& DMoN & 0.79 ($\downarrow 0.09$) & 0.59 ($\downarrow 0.03$) & 0.84 ($\downarrow 0.05$) &  0.47 ($\downarrow 0.03$) \\
\hline
\end{tabular}
\caption{ECS scores for comparison of Nettack and Metattack for community structures with $\rho= 0.1, 0.5$ on ADC-SBM.}
\label{tab:adversarial results_sbm}
\end{table*}


\end{document}